\documentclass[prd,twocolumn,superscriptaddress,letterpaper]{revtex4}
\usepackage{amsfonts}
\usepackage{amsmath}
\usepackage{graphicx}
\usepackage{epstopdf}
\usepackage{natbib}
\usepackage{amssymb}
\usepackage{color}

\renewcommand{\vec}[1]{\mathbf{#1}}
\newcommand{\bm}[1]{{\mbox{\boldmath $#1$}}}

\newcommand{\vs}{\nonumber\\}
\def\ba#1\ea{\begin{align}#1\end{align}}
\newcommand{\be}{\begin{equation}}
\newcommand{\ee}{\end{equation}}
\newcommand{\N}{\mathcal{N}}
\newcommand{\M}{\mathcal{M}}
\newcommand{\refeq}[1]{Eq.~(\ref{eqn:#1})}
\newcommand{\refeqs}[1]{Eqs.~(\ref{eqn:#1})}
\newcommand{\reffig}[1]{Fig.~\ref{fig:#1}}
\newcommand{\refsec}[1]{Sec.~\ref{sec:#1}}
\newcommand{\refapp}[1]{App.~\ref{sec:#1}}
\newcommand{\simgt}{\lower.5ex\hbox{$\; \buildrel > \over \sim \;$}}
\newcommand{\simlt}{\lower.5ex\hbox{$\; \buildrel < \over \sim \;$}}

\def\d{\delta}
\def\s{\sigma}
\def\D{\Delta}

\def\<{\left\langle}
\def\>{\right\rangle}
\def\Mpch{\,h^{-1}{\rm Mpc}}
\def\Msunh{\,h^{-1}M_\odot}

\begin{document}

\title[]{Modeling the Phase-Space Distribution around Massive Halos}

\author{Tsz Yan Lam}
\affiliation{Max Planck Institute for Astrophysics, Karl-Schwarzschild-Str. 1,
85748 Garching, Germany
}
\affiliation{Kavli-IPMU (WPI), the University of Tokyo, Kashiwa-no-Ha 5-1-5, Kashiwa, Chiba 277-8583, Japan}

\author{Fabian Schmidt}
\affiliation{Department of Astrophysical Sciences, Princeton University, Princeton, NJ~08544, USA}
\affiliation{Einstein Fellow}

\author{Takahiro Nishimichi}
\affiliation{Kavli-IPMU (WPI), the University of Tokyo, Kashiwa-no-Ha 5-1-5, Kashiwa, Chiba 277-8583, Japan}
\affiliation{Institut d'Astrophysique de Paris, 98 bis boulevard Arago, 75014 Paris, France}

\author{Masahiro Takada}
\affiliation{Kavli-IPMU (WPI), the University of Tokyo, Kashiwa-no-Ha 5-1-5, Kashiwa, Chiba 277-8583, Japan}

\begin{abstract}
The comparison between dynamical mass and lensing mass provides a
targeted test for a wide range of modified gravity models.  In 
our
previous paper
\citep{vlosDisp12} we showed, through numerical simulations, that the measurement of the line-of-sight
velocity dispersion around stacked massive clusters whose lensing
masses are known allows for stringent constraints on modified gravity on 
scales of $2-15\Mpch$. 
In this work we develop a semi-analytical approach based on the halo
model to describe the phase-space distribution and the line-of-sight
velocity dispersion for different tracers.  The model distinguishes 
contributions from the halo pairwise velocity and the virial
velocity within halos.
We also discuss observational complications, in particular
the contribution from Hubble flow,
and show how our model can incorporate these complications. 
We then incorporate the effects of  modified gravity (specifically, $f(R)$ 
and braneworld models), and show that the model predictions are in excellent agreement 
with modified gravity simulations.  More broadly, the phase-space distribution 
provides a sensitive test of our understanding of hierarchical structure formation when confronted with observations via this model.
\end{abstract}


\maketitle

\section{Introduction}

The $\Lambda$-dominated Cold Dark Matter ($\Lambda$CDM) paradigm, built
on the foundation of Einstein's theory of general relativity (GR), has
proven remarkably successful to explain a broad range of cosmological
observations including the recent Planck measurement of cosmic microwave
background anisotropies \citep{PlanckCosmoParas:13}.  However, the
success has to be recognized as {\em phenomenological} in a sense that the
model requires introducing exotic components of matter and energy, dark
matter and dark energy, where the nature of the dark components has yet
to be known. While the vacuum energy of quantum fields offers a natural
candidate for dark energy, the predicted amplitude from field theory
calculations is many orders of magnitude larger than implied from
observations \citep{Weinberg:89,Friemanetal:08}. Given this fact, there
is growing interest in exploring a possible modification of GR on
cosmological scales as an alternative solution to the dark energy
problem or more precisely the cosmic acceleration problem.

To discriminate dark energy and modified gravity scenarios as the
origin of cosmic acceleration requires one to combine geometrical
probes with large-scale structure probes, where the former constrains
the cosmic expansion history and the latter constrains the growth of
structure formation. Promising probes of large-scale structure are
galaxy clustering \cite{Reidetal:12,Macaulayetal:13}, the abundance
of massive clusters \cite{Vikhlininetal:09,PlanckSZ:13}, and weak
lensing measurements \cite{Heymansetal:13,Addisonetal:13}. 
Ongoing and upcoming wide-area galaxy surveys are aimed at achieving a
higher-precision test of large-scale structure probes; the Baryon
Oscillation Spectroscopic Survey (BOSS)
\footnote{http://www.sdss3.org/surveys/boss.php}, the HETDEX survey
\footnote{http://hetdex.org}, the Extended Baryon Oscillation
Spectroscopic Survey (eBOSS)
\footnote{http://www.sdss3.org/future/eboss.php}, the BigBOSS
\footnote{http://bigboss.lbl.gov/}, the Subaru Prime Focus Spectrograph
(PFS) project \footnote{http://sumire.ipmu.jp/en/2652}
\citep{Ellisetal:12}, the Subaru Hyper Suprime-Cam (HSC) Survey
\footnote{http://www.naoj.org/Projects/HSC/index.html}, the Dark Energy
Survey (DES) \footnote{http://www.darkenergysurvey.org}, the satellite
Euclid mission \footnote{http://sci.esa.int/euclid}, and the LSST
project \footnote{http://www.lsst.org/lsst/}.

It is important to test GR over a wide range of length scales and cosmology
provides many opportunities for such tests, going from the
linear regime to the deeply nonlinear regime, a few 10kpc -- 1Gpc, as
stressed in \cite{JainKhoury:10,Jain:11}.  Clusters of galaxies, the
largest virialized objects in the universe, offer a useful laboratory to
test gravity, because clusters can be studied using various observation
probes such as dynamical probes (velocity field of member and/or
surrounding galaxies), weak/strong gravitational lensing, X-ray and
Sunyaev-Zel'dovich effect (e.g., recall Bullet Cluster as for such a
poster child example to test the nature of dark matter
as demonstrated in \citet{Cloweetal:06}).

Based on this motivation, in 
Paper 1 \citep{vlosDisp12} we proposed a new method of using the
phase-space distribution around massive clusters to probe their dynamical
potential on scales of $2- 15\Mpch$. The phase-space distribution can
be probed by stacking redshift differences of tracer objects (galaxies
or secondary halos) around many massive clusters (halos). Then the
measured dynamical masses can be compared to those
measured from stacked weak lensing
\cite{OguriTakada:11,Okabeetal:13} in order to address whether or not
the dynamical and gravitational masses for the same cluster sample agree
with each other. Thus the method offers a model-independent test of GR
on scales that have not been fully exploited so far.  In Paper 1 we 
demonstrated the ability of the
phase-space distribution function around massive clusters of $\simgt
10^{14}M_\odot/h$ using $N$-body simulations for $\Lambda$CDM and
modified gravity models ($f(R)$ \cite{Caretal03,NojOdi03,Capozziello:2003tk}
and braneworld models \cite{DGP1,DGPMII}).  
In particular we focused on the lowest non-trivial moment of the phase space,
the line-of-sight velocity dispersion,
measured as a function of projected separation between halo pairs.  
We showed that this probe would improve the constraint on modified
gravity parameters by an order of magnitude, if we can use overlapping
imaging and spectroscopic surveys for the same region of the sky that
cover a few thousand square degrees, which is the case for future
surveys such as the Subaru HSC and PFS surveys.
 
The purpose of this paper is to develop a semi-analytical model to
describe the phase-space distribution function around massive clusters,
based on the halo model approach \citep[see][for a review]{haloreview}.  
Since the building blocks of our method are halo-related quantities, we
develop the method in such a way that it is applicable to GR as well as
modified gravity, by including the modifications of the halo abundance
and dynamics in modified gravity, taking into account the non-linear
screening mechanisms present in these models.  
We will also discuss complications arising when applying this method to
actual data, in particular the line-of-velocity contribution due to 
Hubble flow.  We test/calibrate the method by comparing the model predictions 
with $N$-body simulations for GR and modified gravity models. 
We point out that our method is different from methods developed in  
previous studies \citep{tinker07,sd01,cuestaetal08,rw11}.  
The target range of scales for our method is roughly
$2-15 \Mpch$, where we primarily probe the
coherent infall motion of tracer objects towards the central massive halos
(in the following, we will refer to these as the \emph{primary halos}).  
These motions are easier to model theoretically than the random motions within
galaxy clusters on smaller scales, which might be affected by tidal
friction, velocity bias, and baryonic effects.  On the other hand, while
our model describes peculiar motions well on large scales, the upper bound
on the applicability of this method is set by the modeling of the Hubble flow 
contribution, which dominates on scales of $15 \Mpch$ and above.  

We will consider two types of tracers of the phase space: dark matter
particles and intermediate-mass halos (\emph{secondary halos}).  
Actual galaxies can
be seen as somewhat of an intermediate case between these two: while they
are usually physically associated with halos, they also exhibit 
intrahalo motions, which is of course also the case for dark matter itself.  
Recently, \citet{zuweinberg2012} also studied the phase-space structure around
massive halos using $N$-body simulations for the $\Lambda$CDM model.
Our study differs from their work in that our method is
primarily based on an analytical approach, intended to employ as few free
fitting parameters as possible.  Further, we include the extension to modified 
gravity models.

The structure of this paper is as follows. In Section~\ref{sec:nbody},
we briefly describe details of $N$-body simulations to use for testing
our semi-analytical model. In Section~\ref{sec:model}, we develop a
semi-analytic model to describe the stacked phase-space structure around
massive halos, for the halo-halo pairs and halo-dark matter pairs, based
on the halo model approach. In Section~\ref{sec:veldisp}, we study, as
the useful observable of our method, the stacked velocity dispersion of
the tracers measured as a function of the projected radius from the
primary, massive halos. In Section~\ref{sec:comp}, we show the
comparison of the model predictions with $N$-body simulation
measurements, and then discuss the impacts of observational
complications, especially the effect of the Hubble flow, on our method in
Section~\ref{sec:complications}. In Section~\ref{sec:MG}, we
extend our method to modified gravity models, $f(R)$ and DGP
models. Section~\ref{sec:discussion} is devoted to discussion. 
All the detailed calculations and derivations of the model ingredients
are given in Appendix sections.

\section{Phase-space distribution from simulations}
\label{sec:nbody}

\begin{figure}
\begin{center}
\includegraphics[width =0.5\textwidth]{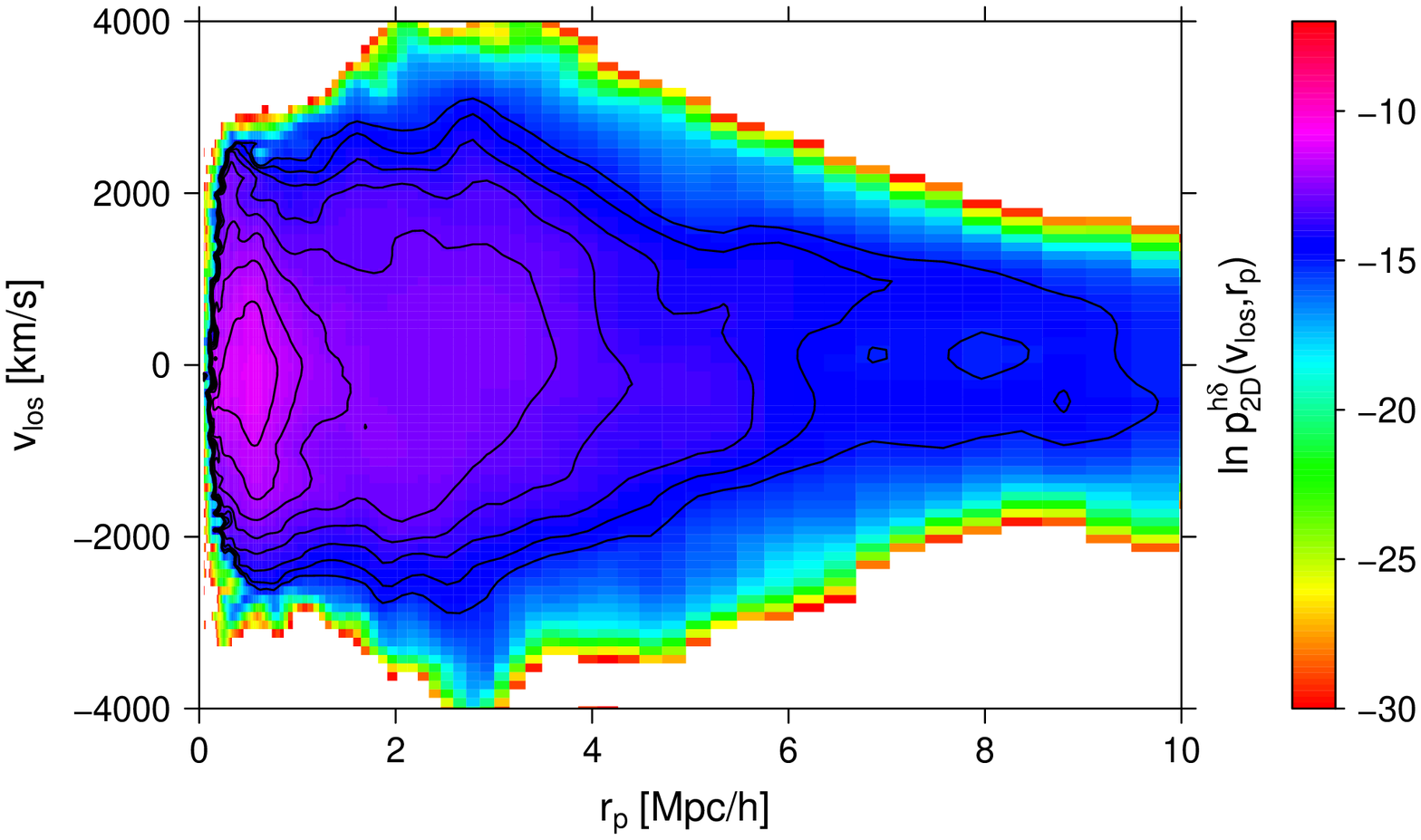}
\includegraphics[width =
0.5\textwidth]{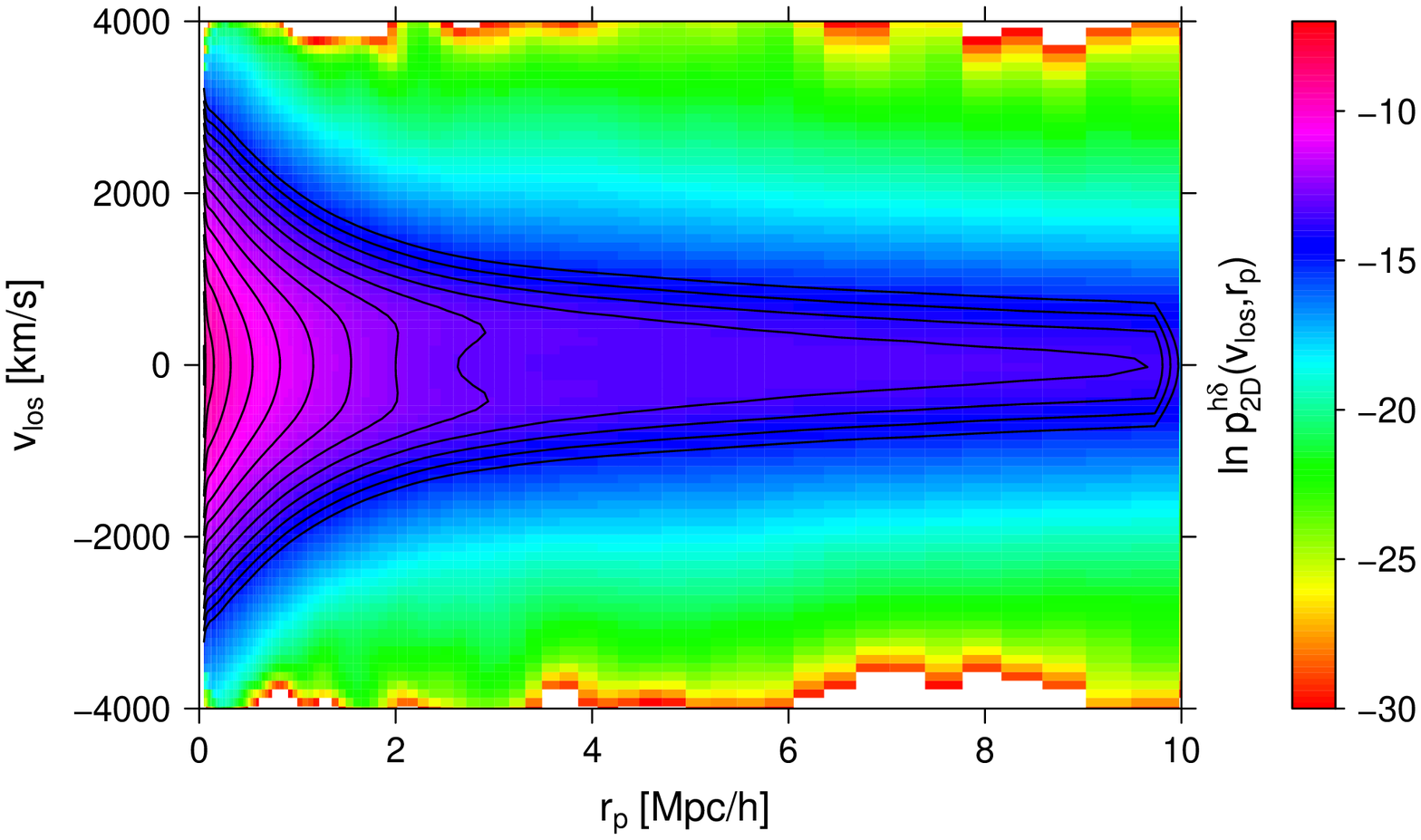}
\includegraphics[width=0.5\textwidth]{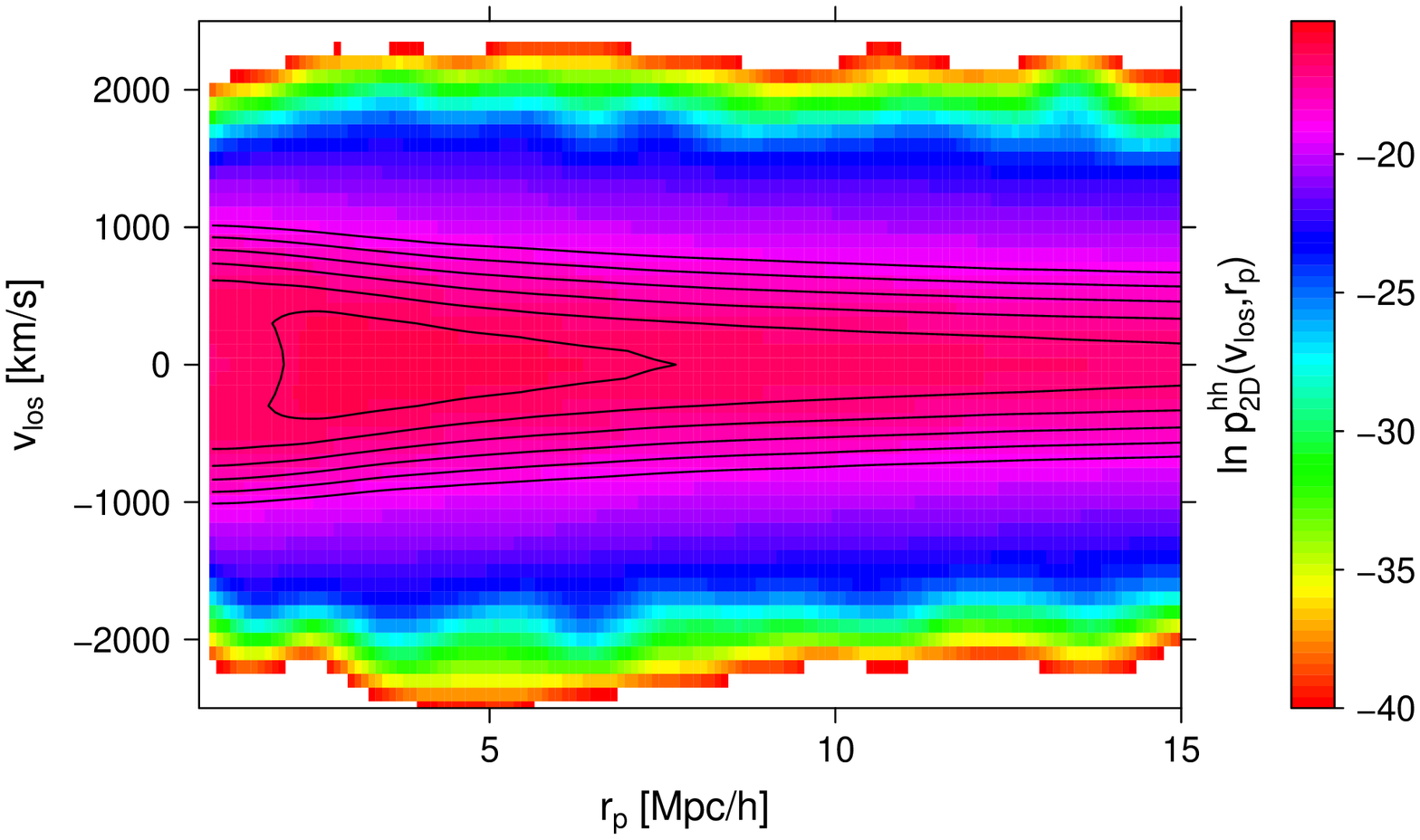}
\end{center}
\caption{The phase-space distribution $p_{\rm 2D}^{h\delta}(v_{\rm vlos},r_p)$ with 
         halo-dark matter particles pairs (upper and middle panels).  
         The upper panel shows the result when only one massive halo 
         ($M=3\times 10^{15}\ M_{\odot}/h$) is used while the middle panel 
         shows the distribution when the most 
         massive 1995 halos 
         ($3\times 10^{14} \leq M_p/M_{\odot}/h < 3\times 10^{15} $) are included.
         Logarithmic scale of the probability 
         is shown in the color scale  and 
         the overplotted contours are the isocontour of the 
         logarithmic of the distribution.  
         The bottom panel shows a similar measurement but using halo pairs 
         and the primary and the secondary halo mass ranges are respectively 
         $10^{14} \leq M_p/M_{\odot}/h < 3\times 10^{15}$ and 
$3\times 10^{13} \leq M_s/M_{\odot}/h < 10^{14}$. 
}
\label{fig:nhalo_1}
\end{figure}

We will present measurements of the phase-space distribution functions around 
primary halos of mass $M_p > 10^{14} \Msunh$ in $N$-body simulations.  
Specifically, we will consider
two tracers, dark matter particles and halos of mass $M_s$ in the range
$3 \times 10^{13} \Msunh \leq M_s \leq 10^{14} \Msunh$.  
For the phase-space distribution of dark matter particles, 
we use one of the realizations described in \citet{vn11_pt_halo} since the
statistics are already sufficient.  This simulation was performed in a box of $1024 \ {\rm Mpc}/h$ on a side with $2048^3$ particles.   
The phase-space distribution of halo-halo pairs was measured from 
a set of 20 realizations performed in a box of $1147.72\ {\rm Mpc}/h$ on a side
with $1280^3$ particles \citep{nt11_halo_halo}, in order to improve statistics.  Both sets of simulations 
adopt the \emph{WMAP} 5-year best-fit parameters in a 
flat $\Lambda$CDM cosmology.  
Measurements were made from the simulation output at $z=0.35$ where
halos are identified by the Friends-of-Friends ({\em FoF}) finder algorithm.  
We measure the relative line-of-sight velocity $v_{\rm los}$ of 
the tracers relative to each primary halo by projecting the relative
velocities of tracers (dark matter particles or halos) with a line-of-sight separation less than $10 \ {\rm Mpc}/h$ ($20 \ {\rm Mpc}/h$) for dark matter particles (halos); see \refsec{complications} for a discussion of this line-of-sight separation cut.
The projected separation $r_p$ is
the distance of the tracers from the halo center on the plane
perpendicular to the line-of-sight direction. 
The phase-space distribution function $p_{\rm 2D}(v_{\rm los}, r_p)$ is
formed by the measured pair of $(v_{\rm los},r_p)$ and we set the 
normalization of the distribution such that 
\begin{equation}
\Delta \ln(r_p) \Delta v_{{\rm los}} \sum_i \sum _j 2\pi {r_{p,i}}^2 p_{\rm 2D}(v_{{\rm
  los},j}, r_{p,i}) = 1.
\end{equation}
In what follows we will use superscripts $h\delta$ and $hh$ to distinguish 
the phase-space distribution functions for halo-dark matter particle pairs
and halo-halo pairs, respectively.  The result is illustrated for
different primary mass ranges and tracers in \reffig{nhalo_1}.  On small
scales $r_p \lesssim 1 \Mpch$, the velocity structure is dominated by
virial motions within the primary halo.  On larger scales, both radial
infall and tangential motions contribute, as well as virial motions within
secondary halos in the case of dark matter.

Dark matter halos are highly aspherical objects and filament
structures are sometimes associated with halos.  The phase-space
distribution measurement for an individual primary halo strongly depends
on the line-of-sight direction as well as the formation/assembly history of
that particular halo.   This complication is alleviated by averaging
(stacking) over many primary halos.  
The effect is illustrated in \reffig{nhalo_1}: the
upper panel shows the dark matter phase-space distribution around the
most massive halo in the simulation ($M_{\rm halo} = 3\times 10^{15}\ M_{\odot}/h$), 
where we averaged over projections along each of the three Cartesian
directions of the simulation box.  The middle panel shows the result by 
stacking the most massive 1995 halos 
(mass range from $3\times 10^{14}$  to $3\times 10^{15}\ M_{\odot}/h$).  We
see that the stacking has produced a smooth distribution symmetric around 
$v_{\rm los} = 0$, as expected since our measurements are made in comoving coordinates.

Finally, the lower panel of \reffig{nhalo_1} shows the phase-space
distribution when using 
secondary halos as the tracers instead of dark matter
(see caption for the mass ranges for the primary and secondary halos.)
The main differences in comparison to the dark matter phase-space are the
lack of virial motion within the primary on small scales (due to the fact
that we do not identify subhalos within the primary);  and a reduced
dispersion at larger $r_p$.  The latter is due to the fact that the 
center-of-mass velocity of secondary halos is obtained by averaging over 
many dark matter particles, thus reducing the dispersion.  Within the 
halo model approach, the dark matter velocities include a component due
to virial motion within the secondary halos, which is absent for the
halos themselves.  We now describe our model for the phase-space distribution
shown in \reffig{nhalo_1}.

\section{Halo model description}
\label{sec:model}

In this section we develop a semi-analytical model based on the halo model
ansatz to describe the phase-space distribution measured in $N$-body simulations.
The ingredients needed to calculate the full phase-space distribution are
the halo-tracer correlation function, the halo-halo velocity distribution and the distribution of virial motion within halos.  
While knowledge of the velocity distributions is essential if we want to calculate the
full phase-space distribution, only moments of these 
distributions are needed to calculate the moments of the phase-space distribution.
For example, the line-of-sight velocity dispersion discussed in Paper 1 
requires only the second moments of these velocity distributions.  

Following the paradigm of the halo model, contributions in the phase-space distribution can be separated into 1-halo and 2-halo terms.  
In this work the 1-halo term describes contributions from tracers lying
within the primary halo while the 2-halo term describes contributions from tracers 
lying outside the primary halo.  We will outline the general building blocks of the calculation here and before moving on to the details in the following subsections.

The phase-space distribution gives the probability of having a line-of-sight 
velocity $(v_{\rm los},v_{\rm los}+{\rm d}v_{\rm los})$ at a projected separation 
$(r_p,r_p + {\rm d}r_p)$,
\begin{align}
p_{\rm 2D}(v_{\rm los},r_p) =\:& \frac1{\N} \int^{M_{p,{\rm max}}}_{M_{p,{\rm min}}} \!\!\!{\rm d}M_p\ 
                 n(M_p) \vs
 \times& \int {\rm d}z\ \rho(r(z,r_p)|M_p)p_{\rm 3D}(v_{\rm
 los}|r,\cos\phi,M_p), 
\label{eqn:phasesp} \\
\N = \int^{M_{p,{\rm max}}}_{M_{p,{\rm min}}} &\!\!\!{\rm d}M_p\ 
      n(M_p) \int\!{\rm d}z\!\int\! {\rm d}r_p 2\pi r_p\rho(r(z,r_p)|M_p),
\label{eqn:NN}
\end{align}
where $r = \sqrt{z^2 + r_p^2}$ is the 3D separation between the primary halo and the 
tracers while $z$ denotes the line-of-sight separation (not
redshift), 
and $\cos\phi = z/r$.  
Here we do not distinguish the tracers but in the following sections 
we will use superscript $^{hh}$ to denotes halo-halo pairs and $^{h\delta}$ 
for halo-dark matter particle pairs.
Note that throughout this paper we assume the distant observer
or plane-parallel approximation.
$p_{\rm 3D}$ is the line-of-sight projected velocity distribution 
for a primary halo-tracer pair separated by $r$, and an angle between 
the separation and the line-of-sight direction $\phi$. 
We assume that $p_{\rm 3D}$ is normalized as
\be
\int d v_{\rm los}\ p_{\rm 3D}(v_{\rm los} | r, \cos\phi, M_p) = 1.
\label{eqn:norm2}
\ee
The first integration in \refeq{phasesp} describes the summation 
(weighted by number density) 
 over primary halo of different masses, 
$M_{p,{\rm max}}$ and $M_{p,{\rm min}}$ being the most and the least massive halos
in consideration and $n(M_p)$ is the halo mass function -- 
this represents the stacking of halos. 
The second integration describes weighted sum 
of the probability of the tracers having 
a relative line-of-sight $v_{\rm los}$ at various line-of-sight separations.  
Here $\rho(z,r_p|M_p)$ is the density of the tracers in
cylindrical coordinate (the azimuth position is suppressed due to spherical 
symmetry) given a halo of mass $M_p$ at the origin.

\subsection{Phase-space distribution for halo-halo pairs}
In this subsection we consider the case where the tracers are dark matter halos.  Since we do not identify subhalos in our simulations, there are no secondary halos within the primary halos.  Thus, the only contribution to the 
phase-space distribution is the 2-halo term, and only the 
halo-halo velocity distribution is needed.  The line-of-sight relative velocity
is given by
\begin{equation}
v_{\rm los} =  \vec{v}_{\rm halo}\cdot \hat{z},
\end{equation}
where $\hat{z}$ denotes the line-of-sight direction and 
$\vec{v}_{\rm halo}$ is the halo-halo pairwise velocity. 
 We can decompose the halo velocity as
\be
\vec{v}_{\rm halo} = 
v_{hr} \hat{\vec{r}} + v_{ht,1} \hat{\vec{e}}_{t1} + v_{ht,2} \hat{\vec{e}}_{t2}, 
\ee
where $\hat{\vec{r}}$ is the unit separation vector connecting
the secondary and primary
halo centers
(such that $v_{hr} < 0$ for infalling motion), and $\hat{\vec{e}}_{t1,2}$ are orthogonal
unit vectors spanning the plane orthogonal to $\hat{\vec{r}}$.  By spherical symmetry,
$v_{ht,1}$ and $v_{ht,2}$ are statistically the same.  We can thus
choose $\hat{\vec{e}}_{t2}$ to be perpendicular to $\hat{z}$, so that 
$\hat{\vec{e}}_{t1}$ lies in the plane spanned by the line of sight and
$\hat{\vec{r}}_p$.  Then, $v_{ht,2}$ does not contribute to the line-of-sight velocity $v_{\rm los}$, and the latter is given by 
\begin{equation}
v_{\rm los} =  v_{hr} \cos\phi + v_{ht}\sin\phi,
\end{equation}
where
we have designated $v_{ht,1}$ 
as $v_{ht}$ since we will not use $v_{ht,2}$ in the following.  

Hence, the velocity probability function in the second integration in \refeq{phasesp} is
\begin{widetext}
\begin{align}
p_{\rm 3D}^{hh}(v_{\rm los}|r,\cos\phi,M_p)  
&=   \int {\rm d}^3\vec{v}_{\rm halo} \ 
          \tilde p_{\rm 3D}^{hh}(\vec{v}_{\rm halo}|r,M_p,M_s)\ 
             \delta_{\rm D}(v_{\rm los} - \vec{v}_{\rm halo}\cdot \hat{z}) \nonumber \\*
 & = \int {\rm d}v_{hr}{\rm d}v_{ht}\ 
     p_{hh}(v_{hr},v_{ht}|r,M_p,M_s)\delta_{\rm D}(v_{\rm los}- v_{hr} \cos\phi - v_{ht}\sin\phi),
\end{align}
\end{widetext}
where the dependence of $\cos\phi$ is solely due to the projection onto 
the line-of-sight direction.

In \refapp{hhpvd} we describe a heuristic approximation 
for $p_{hh}(v_{hr}, v_{ht}|r,M_p,M_s)$, which is given by 
\refeq{linpairwise}.  We assume that the radial and 
tangential components $v_{hr},\,v_{ht}$ of the halo-halo pairwise
velocity are independent, and that each is Gaussian distributed.
Specifically, the variance 
for the radial or tangential component is given, respectively, as
\begin{align}
\sigma_{hr}^2 & =  \left[1 + \frac{2 \beta_{200}}{1+ \xi_{hh}} 
                          - \left(\frac{\beta_{100}}{1+\xi_{hh}}\right)^2\right] 
                           \sigma_{u_{hr}}^2 \label{eqn:vrvar}\\
\sigma_{ht}^2 & = \sigma_{u_{ht}}^2 \label{eqn:vtvar},
\end{align}
where $u_{hr}$ and $u_{ht}$ refer to the statistics for the
linear
velocity difference, 
and the $\beta_{i00}$ are dimensionless coefficients.  
Their definitions are given in \refeq{siguht}.  
Here we have suppressed the dependence on $r$ and $M_p,M_s$ for clarity.

For the radial component, there is also a non-zero mean corresponding to radial infall.  This is evaluated through the spherical collapse model, multiplied by a constant bias to match the prediction of the linear theory in the large-scale limit.  The radial infall from the spherical collapse model around 
a primary halo $M_p$ is (\refeq{vspher})
\begin{equation}
v_{\rm SC}(r|M_p) = -\frac{H(z)}{1+z}r \frac{f(z)}{3}
\delta_c \left[(1+\delta_{\rm NL})^{1/\delta_c} - 1\right],
\end{equation}
where $H(z)$ is the Hubble parameter, $f(z)$ is the linear growth rate, 
$\delta_c$ is the critical density in the spherical collapse model while
$\delta_{\rm NL}$ is the density contrast of the total enclosed mass at 
a distance $r$ from the center of the primary halo. 
This includes both the
primary halo and the associated exterior matter calculated using the 
halo-matter cross-correlation:

\begin{equation}
 M(<r)\equiv \frac{4\pi}{3}r^3\bar{\rho}_m(1+\delta_{\rm NL})
=M_p + M_{\rm shell}(<r), 
\end{equation}
where $M_{\rm shell}(<r)$ is the average mass contribution arising from
the structure around the primary halo, defined as 
$M_{\rm shell}(<r)\equiv \Theta(r-r_{\rm vir})\int_{r_{\rm vir}}^r4\pi r^{\prime
2}dr' \bar{\rho}_m[1+\xi_{h\delta}(r')]$. Here
$\Theta(r-r_{\rm vir})$ is the
Heaviside step function and $\xi_{h\delta}$ is the halo-matter
cross-correlation function given by
\be
\xi_{h\delta}(r) = b(M_p) \xi(r).
\label{eqn:xihm}
\ee
On the other hand, the linear theory prediction is 
\begin{align}
\langle v_{\rm lin,radial} \rangle & = \frac{\beta_{100}}{1 + \xi_{hh}} \sigma_{u_{hr}},
\end{align}
where $\xi_{hh}$ is the halo-halo correlation function and we approximate 
it by the product of the respective linear biases and the dark matter 
correlation function: $\xi_{hh} = b(M_p)b(M_s)\xi(r)$.
Note that $\beta_{100} < 0$ in our convention.  
Our model for the mean radial velocity is then given by matching to linear theory at $r/r_{\rm vir}=r_{20}=20$, hence
\begin{equation}
\langle v_{hr}(r) \rangle = 
      v_{\rm SC}(r) \times 
         \frac{\langle v_{\rm lin,radial}(r_{20}) \rangle}{v_{\rm SC}(r_{20})}.
\label{eqn:vrmean}
\end{equation}
Note that our prescription for the mean infall velocity is the main heuristic
ingredient in the model.  The spherical collapse model by itself does not 
provide a good description of the infall motion as it ignores the angular 
momentum of the infalling matter.  An improved analytical model which
takes the angular momentum into account and matches linear theory on large
scales would be valuable in this context, but we leave this for future work.  

Finally, the density weighting $\rho$ in the second integral of 
\refeq{phasesp} is given by
\begin{equation}
\rho_{\rm halo}(r|M_p) = \int {\rm d}M_s\ n(M_s)
       [1 + \xi_{hh}(r|M_p,M_s)],
\label{eq:rhoh}
\end{equation}
where the integration limit is the mass range of the secondary halo.  
We use the Sheth-Tormen prescription \cite{st02} for the halo mass function $n(M)$.  

Further, we impose halo 
exclusion by setting $\xi_{hh}=-1$ if $r$ is smaller than the sum of the 
virial radii of the two halos.  This exclusion is performed as function
of mass in integrals such as Eq.~(\ref{eq:rhoh}).

\subsection{Phase-space distribution from halo-dark matter pairs}

Within the halo model framework, dark matter particles surrounding the primary halo are categorized into 
two regimes: those  lying within the virial radius of the primary halo 
(1-halo contribution) and those lying outside (2-halo contribution). 
Dark matter particles in the 1-halo regime acquire the virial motion of 
the primary halo; 
those in the 2-halo regime move both with the center of mass of their host 
halo, as described by the model above, as well as within the halo (virial motion).  
Thus,
\begin{equation}
v_{\rm los} = \begin{cases}
           \vec{v}_{\rm vir,p} \cdot \hat{z} & \text{if }  r \leq r_{\rm vir,p}
 \quad \text{(1-halo)}\\
           \vec{v}_{\rm halo}\cdot \hat{z} +  
              \vec{v}_{\rm vir,s} \cdot \hat{z} & \text{if }  r > r_{\rm vir,p} 
  \quad \text{(2-halo)},
  \end{cases}
\end{equation}
where  $v_{\rm vir,p}$ and $v_{\rm vir,s}$ are the virial velocities 
within the primary and secondary halos,  respectively,  and they have
have implicit dependence on the host halo mass.
Note that the virial velocities in general also depend on the 
position within the hosting halo.  

\subsubsection{1-halo regime}
The probability distribution function  
$p_{\rm 3D}^{h\delta}(v_{\rm los}|z,r_p,M)$ and the weighting density function are given by 
\begin{align}
p_{{\rm 3D}, 1h}^{h\delta}(v_{\rm los}|r,\cos\phi,M_p) =\:& 
        \int{\rm d}^3\vec{v}_{\rm vir}\ p_{1h}(v_{\rm vir}|r,M_p) \vs
& \quad \times
              \delta_{\rm D}(v_{\rm los}-\vec{v}_{\rm vir}\cdot \hat{z}), 
                                                 \label{eqn:p1h}  \\
\rho_{\rm DM, 1h}(r| M_p) =\: & \rho_{\rm NFW}(r|M_p). \label{eqn:rho1h}
\end{align}
The subscript $1h$  denotes the 1-halo contribution
 and we omit the superscript $h\d$ since this term does not exist for $hh$.  
The density distribution within halos is described by the NFW profile 
$\rho_{\rm NFW}(r|M)$ \citep{nfw}.  
As described in \refapp{vm}, we assume a Maxwellian velocity distribution for $p_{1h}$, that is a Gaussian in each velocity component with 
1D dispersion related to the virial mass by 
\begin{equation}
\sigma_{\rm DM}(M,z) = \sigma_{\rm DM,
  15}\left[\frac{(1+z)^{3/2}M_{\rm 200b}}{10^{15}M_{\rm
      sun}}\right]^{\alpha},
\label{eqn:sigmaDM}
\end{equation}
where $M_{\rm 200b}$ is the mass of the halo enclosed in a radius containing
a mean density of $200 \bar\rho_m$, and $z$ is the redshift.  Note that we make the
(unrealistic) assumption that the dispersion is constant within the halo.  
For the secondary halos, this assumption only has a small effect.  On the 
other hand, model in detail the phase-space distribution within the primary 
halo is not our aim in this paper.  

Since $\vec{v}_{\rm vir}$ is isotropically distributed,
we let $v_{\rm vir}$ denote solely the line-of-sight projection 
of the virial velocity in the following.

\subsubsection{2-halo regime}\label{sec:2-halo}
The calculation in the 2-halo regime is more involved and 
the separation of the 3D velocity distribution and the density weighting
is not straightforward.  We thus adopt some simplifications as described below.

The velocity and density contributions at some 3D separation $r$ ($> r_{\rm vir}$) 
from the primary halo are coming from 
secondary halos whose centers of mass are located at $\vec{r} +\vec{y}$ 
where $|y|$ is the distance from the center of the secondary halo. 
Hence 
\begin{widetext}
\begin{eqnarray}
\rho_{\rm DM, 2h}(r| M_p)  p_{\rm 3D, 2h}^{h\delta}(v_{\rm los}|r,\cos\phi,M_p) &=
      \int{\rm d}M_s \  n(M_s)\int{\rm d}^3 \vec{y}\ 
            \int{\rm d}^3\vec{v}_{\rm halo} 
               \int {\rm d}v_{\rm vir} 
                  [1+\xi_{hh}(\vec{r}+\vec{y}|M_p,M_s)]\rho_{\rm NFW}(y|M_s)
                       \nonumber \\
           & \times p_{2h}(\vec{v}_{\rm halo},v_{\rm vir}||\vec{r}+\vec{y}|,M_p,M_s)
                      \delta_{\rm D}[v_{\rm los} -(\vec{v}_{\rm halo}+
                              \vec{v}_{\rm vir})\cdot \hat{z}],
\label{eqn:rhop2h}
\end{eqnarray}
\end{widetext}
where $M_s$ is the mass of the secondary halo residing at $\vec{r}+\vec{y}$.
The subscript $2h$ denotes the 2-halo contribution.  Again, we assume that 
the virial velocity does not depend on the distance $y$ from the halo center,
so that $p_{2h}$ does not explicitly depend on $y$.  

The 2-halo term as given in \refeq{rhop2h} is computationally expensive to evaluate.  Since we are interested in the scales $r$ significantly larger than the virial radii of the primary halos, several approximations can be made to significantly simplify  \refeq{rhop2h} (for a detailed description see 
\refapp{Qvir}):
\begin{enumerate}
\item The virial motion within the secondary halo is assumed independent 
      of the peculiar motion of the secondary halo.  
\item We neglect the dependence on $M_s$ in the halo-halo pairwise velocity 
      distribution (see \refapp{hhpvd}).
\item We approximate $\vec{r}+\vec{y} \approx \vec{r}$ in the halo-halo clustering
      since $y$ is of order the virial radius of secondary halos.
\item The virial velocity dispersion is assumed to scale as $M^{1/3}$ and 
      the linear density power spectrum is approximated as power-law $\propto k^{-1}$.
\item We set $b(M_s) = 1$ (see \refapp{Qvir} for the alternative expressions without 
       this assumption).
\end{enumerate}
We emphasize that we make these approximations for computational convenience,
since we have found that they do not significantly degrade the accuracy of the model.  With these approximations, \refeq{rhop2h} simplifies to (in analogy with \refeqs{p1h}~and~\eqref{eqn:rho1h}) 
\begin{align}
p_{\rm 3D, 2h}^{h\delta}(v_{\rm los}|r,\cos\phi,M_p) =\:&
     \int {\rm d}v_{hr}{\rm d}v_{ht}{\rm d}v_{\rm vir,s} \vs
 \times
       p_{hh}(v_{hr},v_{ht}|&r,M_p) q_{\rm vir}^{\rm ST}(v_{\rm vir,s}|r,M_p) \vs
\times \delta_{\rm D}(v_{\rm los} -  v_{hr} &\cos\phi - v_{ht}\sin\phi - v_{\rm vir,s}), \vs
\rho_{\rm DM, 2h}(r|M_p) =\:& \left[1+\xi_{h\delta}(r|M_p)\right]\bar{\rho}_m,
\end{align}
where $q_{\rm vir}^{\rm ST}$ is the mass-weighted probability density 
function for the virial velocity assuming a Sheth-Tormen mass function 
(\refapp{Qvir})
\begin{align}
q^{\rm ST}_{\rm vir} (v|r,M_p) =\:& \frac{A(p)}{\pi}\frac{\sqrt{q}}{\sigma_{\rm eff}(r, M_p)} \vs
\times 
  \Bigg[  K_0 &\left(\frac{\sqrt{q}}{\sigma_{\rm eff}}|v|\right) 
            + \left(\frac{\sqrt{q}}{\sigma_{\rm eff}}|v|\right)^{-p}
                  K_p\left(\frac{\sqrt{q}}{\sigma_{\rm eff}}|v|\right)
     \Bigg],
\nonumber
\end{align}
where $A(p)$, $p$, and $q$ are parameters in the Sheth-Tormen mass function.
$\sigma_{\rm eff}(r,M_p)$ is the effective virial velocity dispersion 
of the secondary halo which is described in
\refapp{effsigma}.  In order to improve the accuracy, we in fact only
apply the first two simplifications in the list above to obtain $\sigma_{\rm eff}$.  It is given by 
\begin{align}
\sigma_{\rm eff}^2(r,M_p)  =\:& \frac{1}{\bar{w}}\int {\rm d} M_s
       \int_0^{r_{\rm vir, s}} {\rm d}^3 y\ n(M_s) \rho_{\rm NFW}(y) \vs
& \times            \left[1 + \xi_{hh}(|\vec{r}+\vec{y}|,M_p,M_s)\right] \sigma_{\rm DM}^2(M_s),
\label{eqn:sigeff1}
\end{align}
where $\bar{w}$ is a normalization factor.
This integral can be tabulated as function of $r$ for each primary halo 
mass.  We have found that further approximations to $\sigma_{\rm eff}$ 
significantly sacrifice the accuracy of the model.  See \refapp{effsigma} 
for details.

\subsection{Velocity dispersion of the phase-space distribution}
\label{sec:veldisp}

The previous section describes how to use the halo model to compute the
phase-space distribution around massive halos.  The distribution is
symmetric around $v_{\rm los} = 0$, so that all odd moments of the distribution
vanish.  Thus, the lowest moment that contains information is the dispersion
$\sigma_{v_{\rm los}}$, given in terms of the phase-space distribution by
\begin{equation}
\sigma^2_{v_{\rm los}}(r_p) = \int {\rm d}v_{\rm los}\ v_{\rm los}^2 \:p_{\rm 2D}(v_{\rm los}|r_p)\,,
\label{eqn:sigma2}
\end{equation}
 where $p_{\rm 2D}(v_{\rm los}|r_p)$ is the \emph{normalized} line-of-sight velocity distribution at projected radius $r_p$.  That, it is given by Eq.~(\ref{eqn:phasesp}) but with a different normalization $\N\to \N'(r_p)$, 
\be
\N'(r_p) =\:  \int^{M_{p,{\rm max}}}_{M_{p,{\rm min}}} \!\!\!{\rm d}M_p\ 
      n(M_p) \int {\rm d}z\ \rho(r(z,r_p)|M_p)\,.
\label{eqn:pvz}
\ee
The quantity $\sigma^2_{v_{\rm los}}(r_p)$ was proposed as sensitive test of
modified gravity in Paper 1.  Instead of evaluating \refeq{sigma2} directly,
we instead use the characteristic function of the velocity distribution 
\refeq{phasesp} to compute this quantity, which is simpler and faster. 
We will leave the details of derivation in Appendix~\ref{sec:Qvir} 
and only give the results here.

The line-of-sight velocity dispersion of for halo-dark matter is
\begin{equation}
\sigma_{v_{\rm los}, {\rm DM}}^2 = \frac{\int{\rm d}M_p\ n(M_p)\int {\rm d}z\ \mathcal{F}_{\rm DM}(z,r_p,M_p)}{\int{\rm d}M_p\ n(M_p)\int {\rm d}z\ \rho_{\rm DM}(z,r_p,M_p)},
\label{eqn:DMdisp}
\end{equation}
where $\mathcal{F}_{\rm DM}$ denotes the mass-weighted dark matter velocity
dispersion given by
\begin{widetext}
\begin{equation}
\mathcal{F}_{\rm DM}(z,r_p,M_p) = \begin{cases}
                \rho_{\rm NFW}(r|M_p) \sigma_{\rm DM}^2(M_p) & \text{(1-halo)}\\
                \bar{\rho}_m [1+\xi_{h\delta}(r|M_p)] 
              \left[(\sigma_{hr}^2+ \langle v_{hr}\rangle^2)\cos^2\phi + \sigma_{ht}^2\sin^2\phi  
 - \mathcal{Q}_{\rm vir}^{{\rm ST} \, ''} (t=0)\right]& \text{(2-halo)}\\
               \end{cases} 
\,.
\label{eqn:sigDM}
 \end{equation}
The various terms in the last square brackets in the 2-halo term of 
 $\mathcal{F}_{\rm DM}$ are the different contributions to 
the line-of-sight velocity dispersion, they are respectively 
the halo-halo radial velocity dispersion, the halo-halo radial infall velocity,
the halo-halo tangential velocity dispersion, and the
velocity dispersion due to the virial motion within secondary halos.  
Note that we have assumed that the halo radial and tangential velocities
are statistically independent.

Dropping the 1-halo term as well as the virial velocity contribution in the 
secondary halo, 
the line-of-sight velocity dispersion for halo-halo phase-space distribution is
consequently
\begin{equation}
\sigma_{v_{\rm los}, {\rm halo}}^2 = \frac{\int{\rm d}M_p\ n(M_p)\int {\rm d}z\ 
                [(\sigma_{hr}^2 + \langle v_{hr}\rangle^2)\cos^2\phi + \sigma_{ht}^2\sin^2\phi] 
               \int {\rm d}M_s\ n(M_s)[1+\xi_{hh}(r|M_p,M_s)]
            }{\int{\rm d}M_p\ n(M_p)\int {\rm d}z\,
              \int{\rm d}M_s n(M_s)[1+\xi_{hh}(r|M_p,M_s)]}  
\label{eqn:sighalo}
\end{equation}
\end{widetext}
where again halo exclusion is enforced by $\xi_{hh}(r|M_p,M_s)= -1$ when
$r < r_{\rm vir}(M_p) +  r_{\rm vir}(M_s)$.

\section{Comparison with simulations}
\label{sec:comp}

In this section we compare the predictions of the model described in 
the previous section to measurements from $N$-body simulations.
We compare the 2D phase-space distribution, the projected line-of-sight velocity
distribution, as well as its dispersion.  We first present the result for halo-dark matter particle pairs and then halo-halo pairs.

\subsection{Halo-dark matter distribution}
\label{sec:comphdm}

\begin{figure}
\centering
\includegraphics[width =
0.5\textwidth]{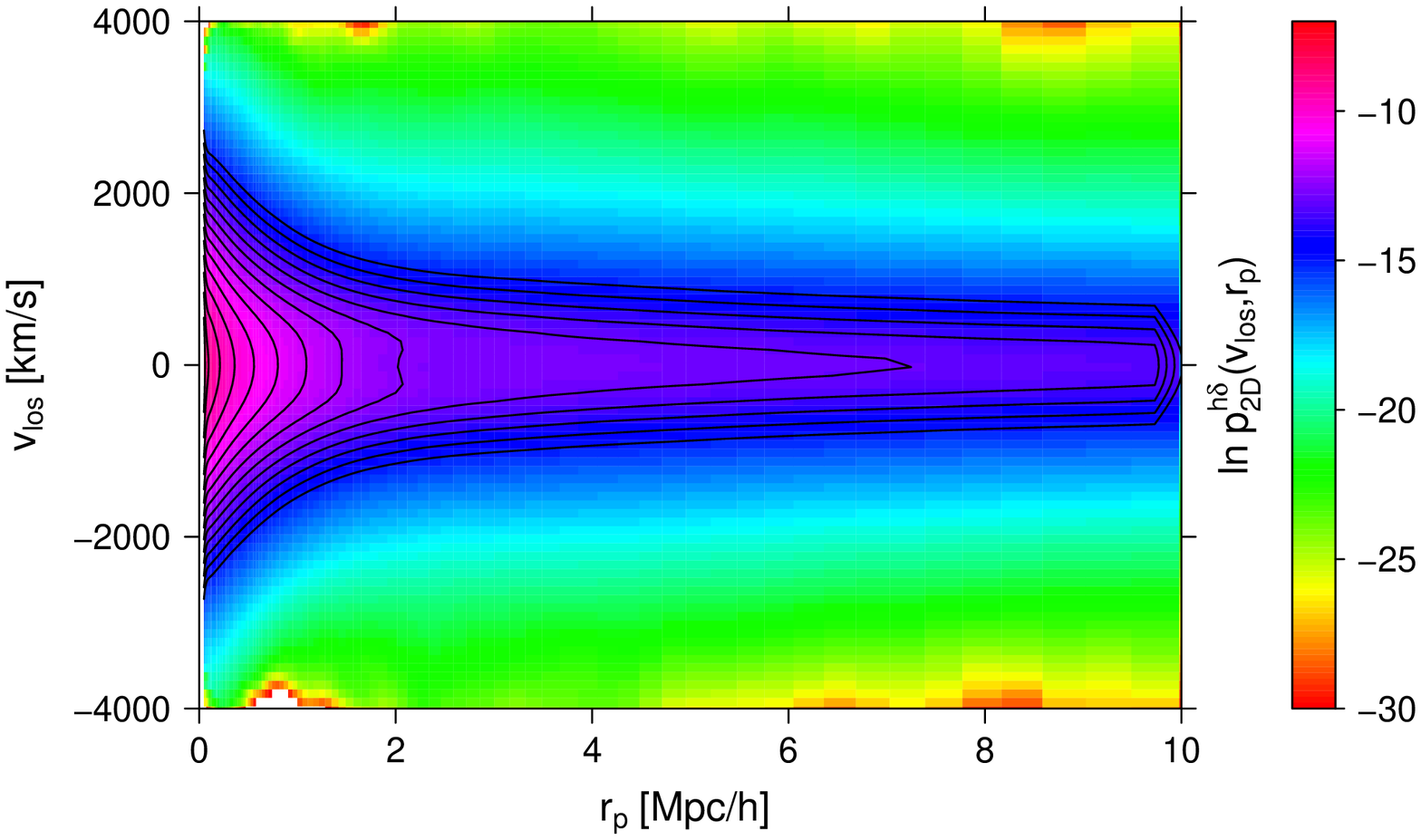}
\includegraphics[width =
0.5\textwidth]{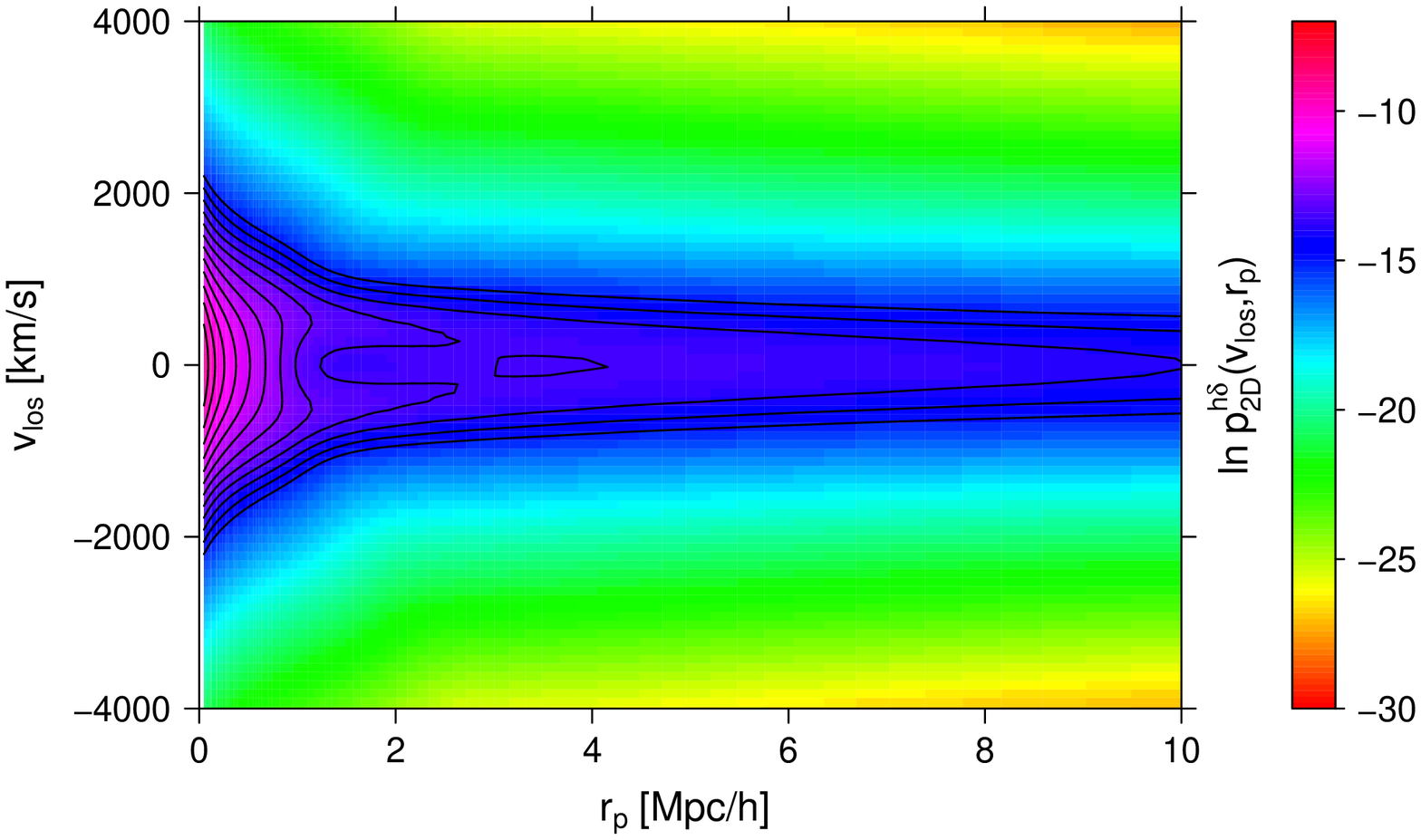}
\caption{The phase-space distribution $p_{2d}^{h\delta}(v_{\rm vlos},r_p)$ in
  logarithmic scale including all the dark matter particles
  surrounding halos more massive than $1\times 10^{14}\ M_{\rm
    sun}/h$. The top panel shows the measurement from numerical
simulation and the bottom panel shows the empirical model (using 
\textit{halofit} nonlinear matter power spectrum) prediction.
The contours are the isocontour of $\ln(p)$, starting from $-15$ with
$+0.5$ increment inwards.}
\label{fig:lnvlosr2d_1e+14_ronrvir_0}
\end{figure}

We first study the phase-space distribution $p_{\rm 2D}^{h\delta}(v_{\rm los},r_p)$ of dark matter
particles around primary halos with $M_p \geq 10^{14}\Msunh$, normalized to
over $v_{\rm los}$ and $r_p$ [\refeqs{phasesp}--\eqref{eqn:NN}].  
\reffig{lnvlosr2d_1e+14_ronrvir_0} shows such a comparison, where the
simulation measurements are shown in the top panel and the model in the
bottom panel.  The color scale and contours correspond to the phase-space
density in logarithmic scale; the scale is the same for both panels.  

Overall the model prediction matches the $N$-body measurement well.  
The most noticeable difference between the simulation and model 
occurs at the transition from 1-halo to 2-halo regime,
around $r_p\approx 2\ {\rm Mpc}/h$.  
In our model (as in most halo model calculations) there is a sharp transition 
at the virial radius of the primary halos.  On the other hand, the actual
velocity and density distributions of dark matter particles do not show such 
a sudden transition.

\begin{figure}
\centering
\includegraphics[width=0.48\textwidth]{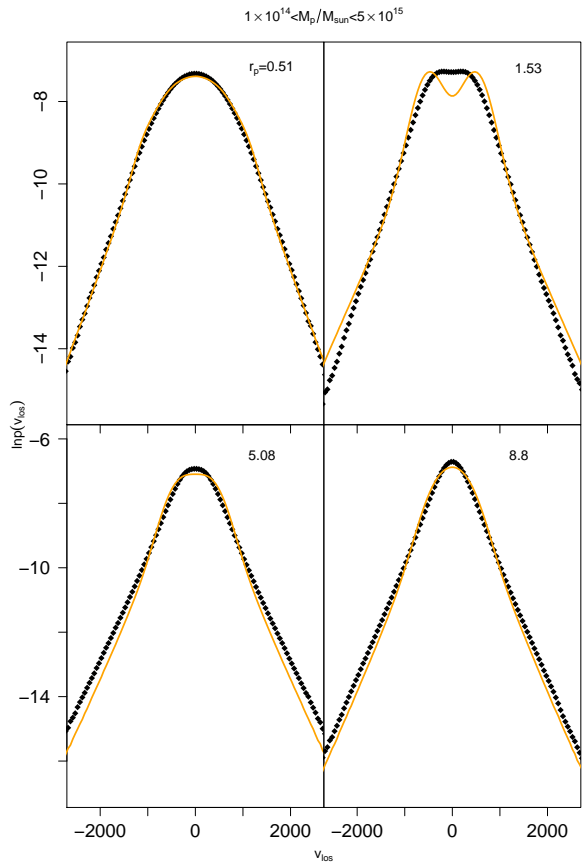}
\caption{The line-of-sight velocity probability distribution 
         at different projected radii, normalized to unity at each $r_p$. 
         The black symbols are 
         $N$-body measurements while the solid orange curves are the model 
         predictions following \refsec{model}.
\label{fig:stats_1e14}}
\end{figure}

\begin{figure}
\centering \includegraphics[width=0.48\textwidth]{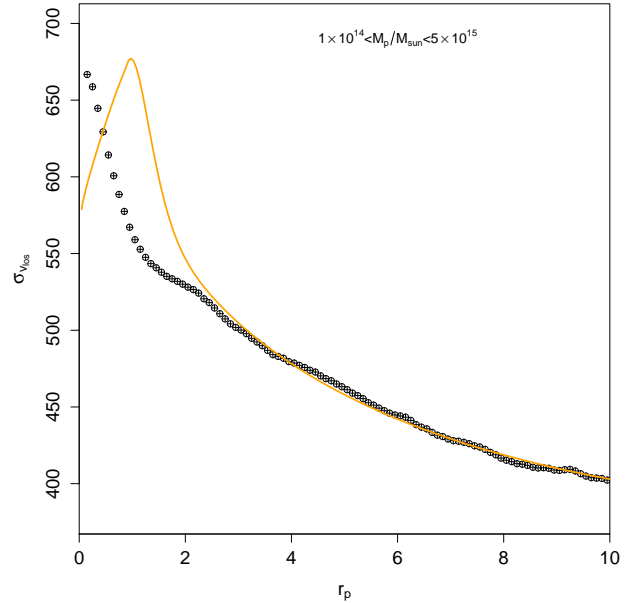}
\caption{Line-of-sight velocity dispersion $\s_{v_{\rm los}}$ (in km/s) as function of projected radius $r_p$.  Solid (orange) curves are predictions from the model as described in \refsec{veldisp}.   }
\label{fig:stats_1e14b}
\end{figure}

\reffig{stats_1e14} shows the distribution of the line-of-sight velocity
at different projected separations $r_p$, while \reffig{stats_1e14b} shows the dispersion (second moment) as function of $r_p$.  For these figures, the distribution is normalized in $v_{\rm los}$ at each $r_p$ following \refeq{pvz}.  
The black symbols are measurements from numerical simulations and 
the orange curves are predictions of the model described
in the previous section.  The central part of the distributions shown in 
\reffig{stats_1e14} is generally
well matched by the model apart from the transition region $r_p \sim 2 \Mpch$.  
This is even more evident in \reffig{stats_1e14b}, showing an excellent prediction for $\s_{v_{\rm los}}$ for $r_p \gtrsim 2.5 \Mpch$.  
At larger separations, the tails in the $v_{\rm los}$ distribution are underpredicted by the model.  This is confirmed by a comparison of the kurtosis of $v_{\rm los}$ in simulations as compared to the model, which underpredicts the kurtosis by about a factor of two.  This departure is not surprising since we have assumed Gaussian velocity distributions throughout, while dynamical processes are generally expected to produce non-Gaussian velocity distributions.

\begin{figure}
\centering
\includegraphics[width=0.5\textwidth]{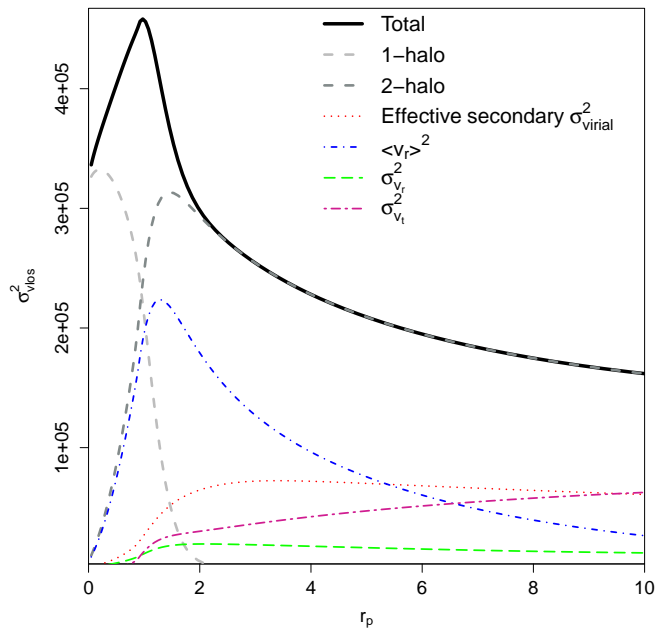}
\caption{Different components contribution to the line-of-sight velocity
dispersion squared for halo-dark matter pairs (see \refeq{sigDM}).  
} \label{fig:sigvlos_GR_split}
\end{figure}

It is instructive to break down $\s^2_{v_{\rm los}}$ into individual contributions.  This is shown in \reffig{sigvlos_GR_split}.  The peak near $r_p=2\ {\rm Mpc}/h$ which is responsible for the main discrepancy with the simulation results is 
caused by the mean radial infall contribution (blue dash-dotted).  At scales approaching the virial radius of the main halo, the spherical collapse model is not a good approximation anymore even when matched to linear theory on large scales.  This is presumably again caused by not taking into account the angular momentum, which leads to a significant overprediction of the radial velocity on small scales.  We have also studied the impact of changing the density profile in the transition region, by extrapolating the NFW profile $\propto r^{-3}$ of the primary halo instead of truncating it at the virial radius.  However, this significantly worsened the model agreement for larger $r_p$.

The model also fails in the innermost region, dominated by the 1-halo contribution.  This probably indicates that our treatment of constant velocity dispersion
within the primary halo is not accurate.  However, as pointed out above, we are not primarily interested in modeling the velocity distribution within the primary halo.

\subsection{Halo-halo distribution}
\label{sec:comphh}

\begin{figure}
\centering
\includegraphics[width =
0.5\textwidth]{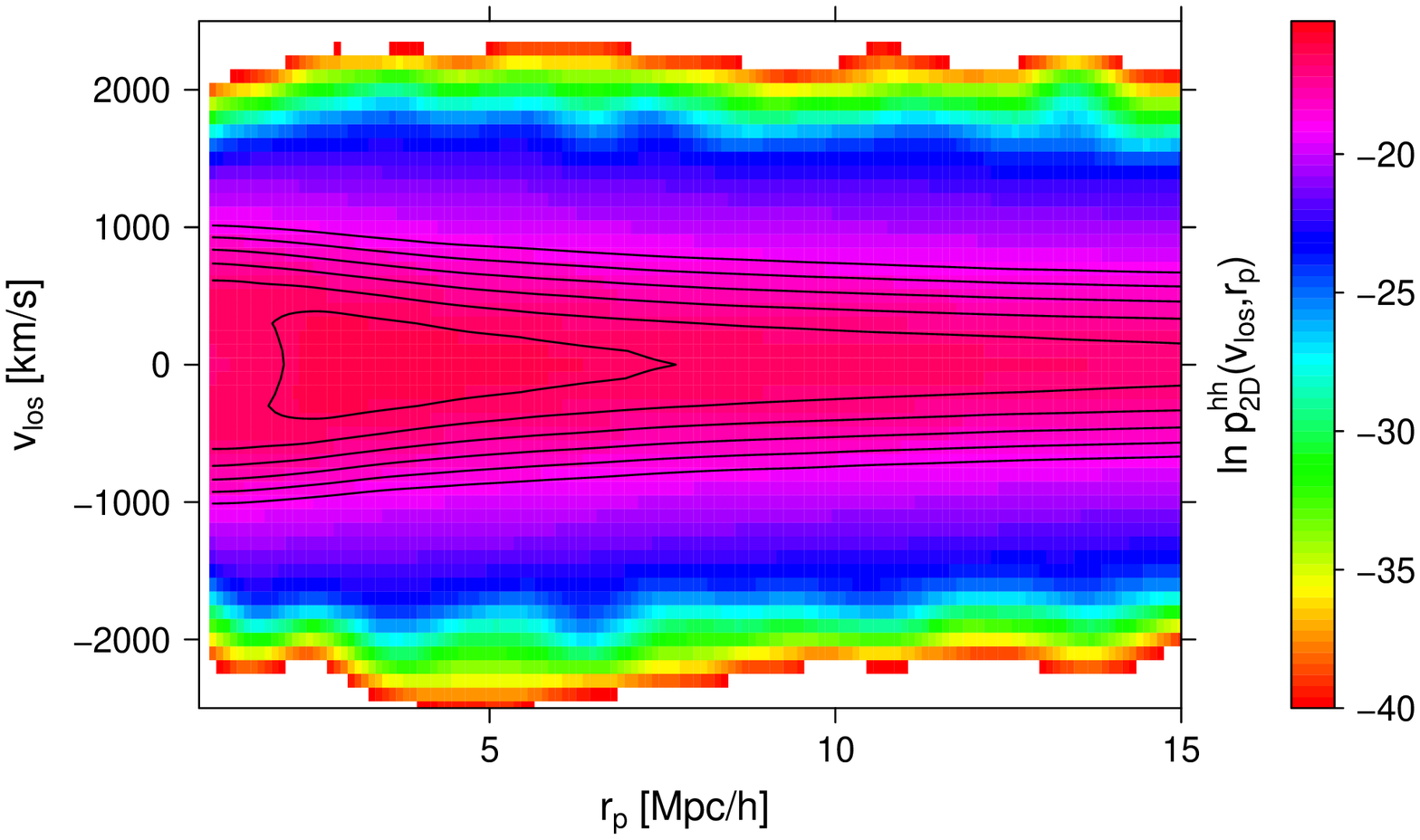}
\includegraphics[width =
0.5\textwidth]{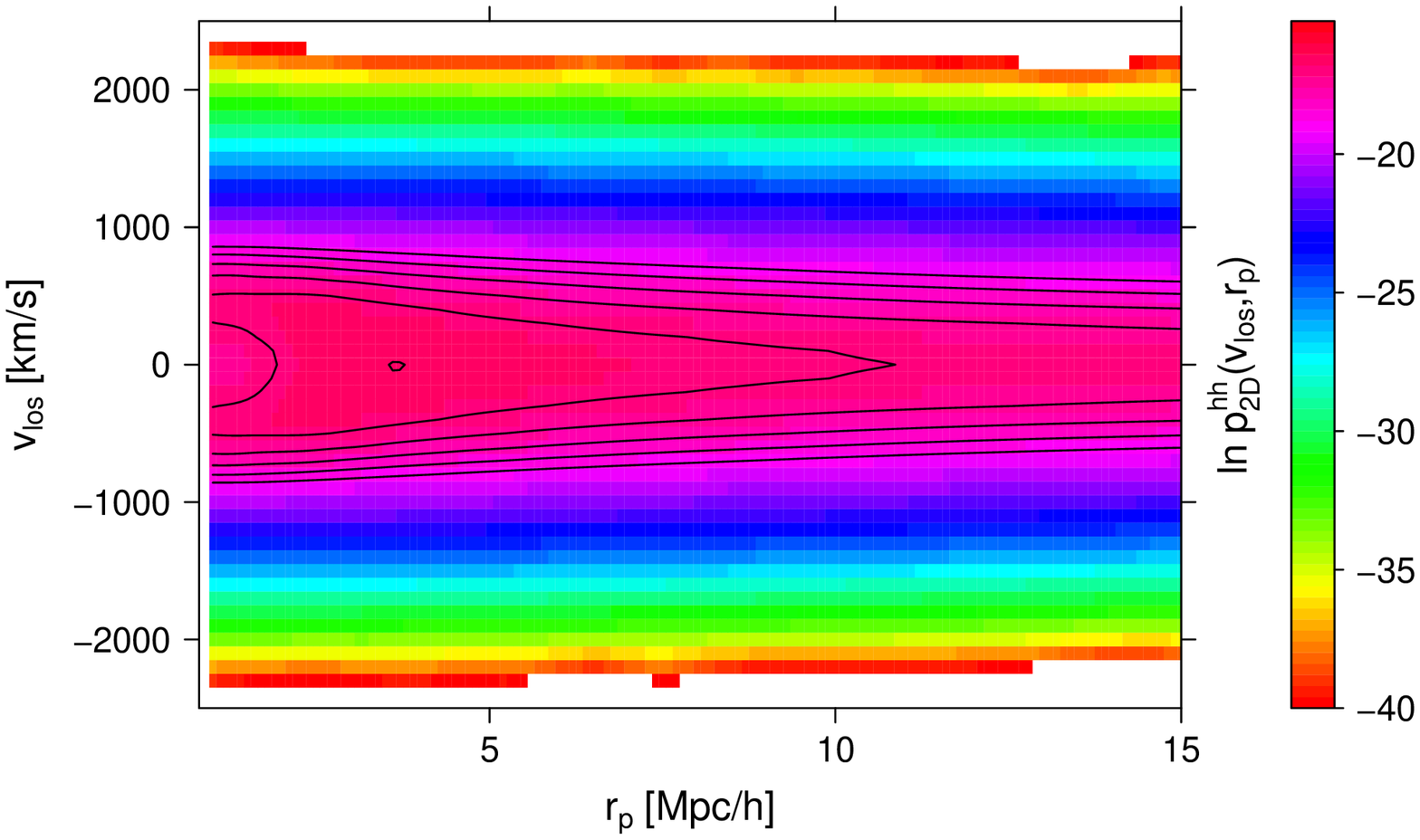}
\caption{The phase-space distribution $p_{2d}^{hh}(v_{\rm vlos},r_p)$ in
  logarithmic scale for halos of mass $3\times 10^{13} \leq M_s/ M_{\odot} < 1\times 10^{14}$ around primary halos with $M_p \geq 10^{14} \Msunh$.  
   The top panel shows the $N$-body measurements while the bottom panel shows the empirical model prediction, where we have used the halo-halo correlation 
function from simulations.  
The contours are the isocontour of $\ln(p)$, starting from $-19$ with
$+0.5$ increment inwards.  The simulation results are calculated for linear
bins in $r_p$ with width $\D r_p = 1\Mpch$.  The first column of data corresponds
to the bin $0 \leq r_p \leq 1\Mpch$ and is shown at $r_p=0.5\Mpch$.
}
\label{fig:lnvlos_halo}
\end{figure}

We now turn to the phase-space distribution of secondary halos
with $3\times 10^{13} \leq M_s/ M_{\odot} < 1\times 10^{14}$, again
around primary halos with $M_p \geq 10^{14} \Msunh$.  
\reffig{lnvlos_halo} shows the distribution in the simulations (top panel)
and the model prediction (lower panel).  Clearly, the model
describes the qualitative features of the phase-space distribution for
halos as well.  Note the overall smaller dispersion and the absence of
the primary halo virial motions.

\begin{figure}
\centering
\includegraphics[width=0.48\textwidth, clip=true, trim=0.75cm 1.5cm 2.2cm 0.5cm]{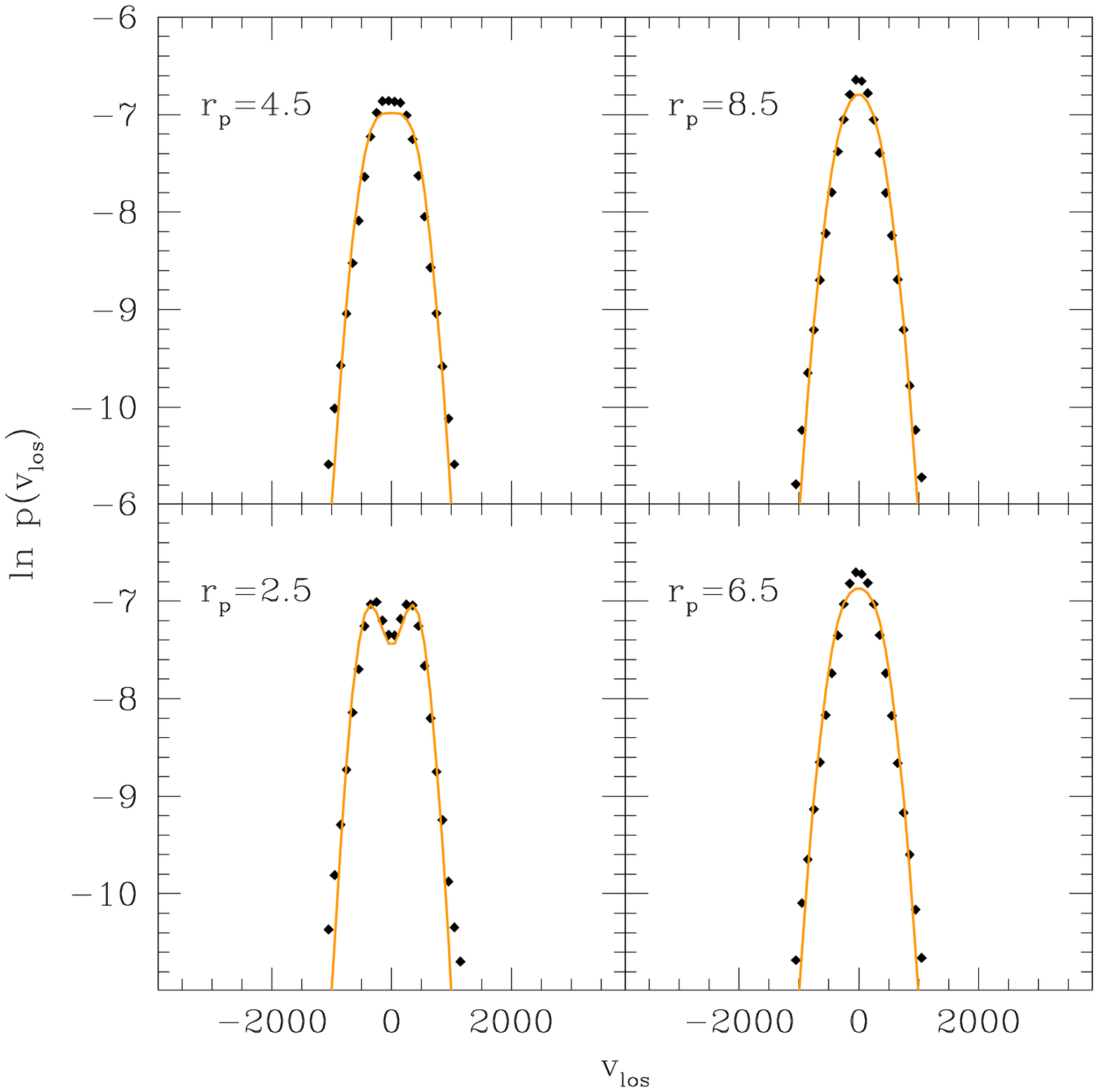}
\caption{The halo-halo line-of-sight velocity distribution at different projected radii, in analogy to \reffig{stats_1e14}.  Black points denote the simulation measurements, orange curves are the model prediction. }
\label{fig:p2d_halo}
\end{figure}

The distribution of the line-of-sight velocity at different projected 
separations is shown in \reffig{p2d_halo} (normalized to unity), 
where again black symbols show
the simulation results whereas the model is shown as orange solid. 
The model predictions match the measured profiles well, including the double peak
feature at small $r_p$ (lower left panel).  The dispersion $\s_{v_{\rm los}}$
is shown as a function of $r_p$ in \reffig{hubble_sharpcut} (lowest set of
points, simulations, and black solid line, model), showing excellent agreement 
over the entire range of scales considered.

\section{Connecting to observations}\label{sec:complications}

The comparisons in the previous section demonstrate that our halo model
based approach can describe the phase-space distribution around massive
primary halos for both dark matter and secondary halos.  However, in order to model
actual data, we need to take into account that peculiar velocities
are not measured directly, but rather through the observed redshift which
is also impacted by the Hubble flow.  In this section we attempt to include 
this effect in the model.  We will illustrate the ideas for the halo-halo pairs, since this case is more closely related to actual galaxy surveys.

The redshift receives contributions both from peculiar motions and the
cosmological redshift.  The observationally inferred line-of-sight velocity difference between the primary halo and tracer is then given by
\be
v_{\rm los,obs} = v_{\rm los} + H z,
\label{eqn:vhubble}
\ee
where $H$ is the Hubble parameter evaluated at the redshift of the primary
halo, and $z$ is the line-of-sight separation of primary halo and tracer
as before.  Here we have assumed the plane-parallel approximation and that positive $v_{\rm los}$ corresponds to
motion away from the observer.  Given the typical velocities in the 
halo phase space (\reffig{p2d_halo}), this effect becomes important when
\be
H z \sim v_{\rm los} \sim 300 \,{\rm km}/{\rm s}\,,
\ee
corresponding to $z \sim 3 \Mpch$.  Thus, for halo pairs with line-of-sight
separation greater than a few Mpc, the Hubble flow contribution cannot be
neglected.  Furthermore, in our analysis in \refsec{model} we have imposed
a cut on the line-of-sight separation between primary and secondary halos,
which cannot be imposed in reality.  

\begin{figure}
\centering
\includegraphics[width=0.48\textwidth, clip=true, trim=0.75cm 1.5cm 2.2cm 0.5cm]{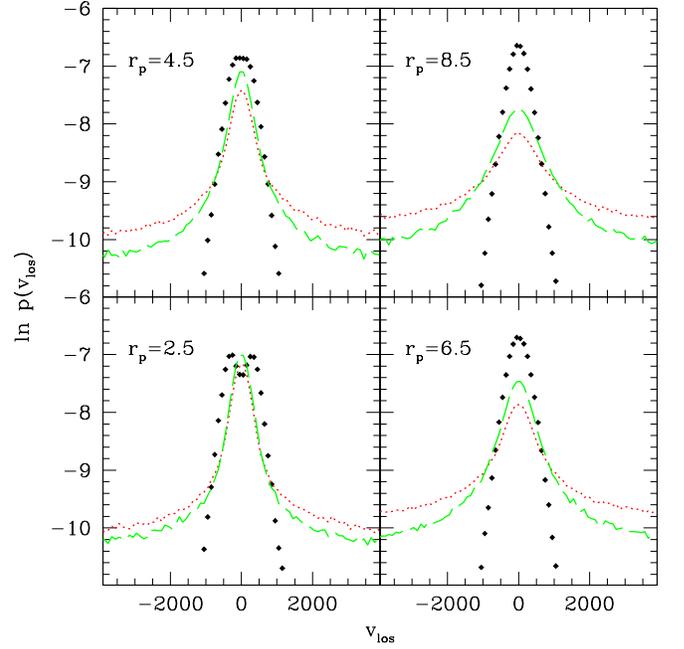}
\caption{The halo-halo line-of-sight velocity distribution at different projected radii:
black points are the distributions for halo pairs selected by their 
line-of-sight separation and without Hubble flow contribution;
red curves are distributions that includes Hubble flow; 
green are the distributions after subtracting a constant interloper contribution 
 (see text). 
}
\label{fig:p2d_vcut}
\end{figure}

\begin{figure}
\centering
\includegraphics[width =0.48\textwidth, clip=true, trim=0.5cm 1cm 1.9cm 1cm]{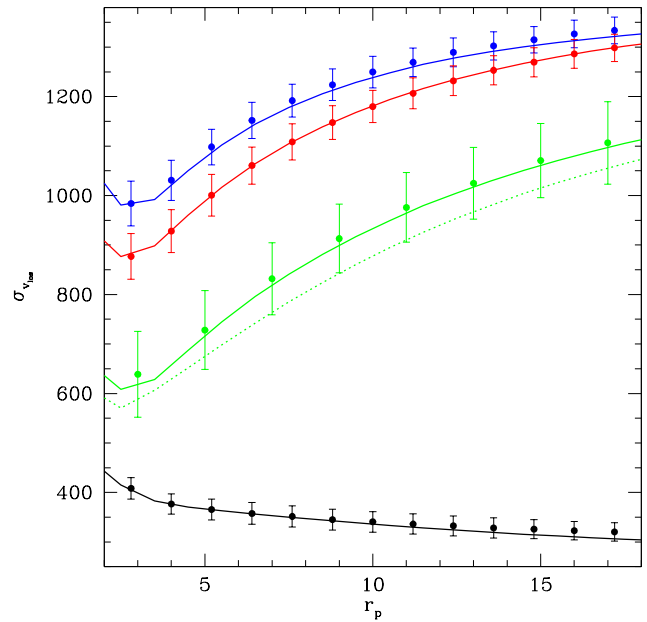}
\caption{ 
The line-of-sight velocity dispersion from halo-halo pairs 
including the Hubble flow.  The different sets of curves and symbols are (from top to bottom):
pure Hubble flow (no peculiar velocities; blue);
Hubble flow and peculiar velocities (red); 
Hubble flow and peculiar velocities after subtracting a constant 
interloper contribution (see text); 
\textit{no} Hubble flow and halo pairs with 
a fixed line-of-sight separation cut 
($|\Delta r_{\rm los}|\leq 20\ {\rm Mpc}/h$, in black).  In the first three
cases, the dispersion is calculated for $|v_{\rm los}| < v_{\rm cut}=2372\ {\rm km/s}$.  
The dotted green curve shows the interloper-subtracted dispersion when 
the halofit correlation function is used instead of the one measured in the
simulations.}
\label{fig:hubble_sharpcut}
\end{figure}

In order to study these effects, we repeat the simulation measurements, but instead of imposing 
a fixed small line-of-sight separation boundary around the primary halo, 
we include all halo pairs within $100\Mpch$ line-of-sight separation
(safely including all pairs that contribute to the range in $v_{\rm los}$ 
we are interested in), and measure the relative line-of-sight velocity 
between halo pairs including the difference in Hubble flow.  
\reffig{p2d_vcut} illustrates these effects in terms of the line-of-sight velocity distributions at 
different projected radii.  The black solid symbols are without the Hubble
contribution, as shown in \reffig{p2d_halo}.  The dot-dashed 
curves show the realistic case when including the Hubble flow and no line-of-sight separation restriction is imposed.  Note the long tails of the distribution due to pairs at large line-of-sight separations leading to correspondingly large $v_{\rm los,obs}$.  At very large $v_{\rm los,obs}$, the contribution is entirely dominated by the Hubble flow, and asymptotes to a constant as the density of tracers approaches its mean (that is when $\xi_{hh}(r)$ becomes negligible).  Clearly, velocity dispersion is not strictly defined in this case.  \reffig{hubble_sharpcut} instead shows the dispersion measured within $|v_{\rm los,obs}| \leq 2372 {\rm km/s}$ (see below) as red points.

Observationally, this constant ``interloper'' contribution will be
subtracted by measuring the constant to which the PDF asymptotes at very large 
values of $v_{\rm los,obs}$, and subtracting this constant (in reality, effects
such as the redshift-dependence of the selection function of the sample lead
to a interloper contribution that is in fact slowly varying with $v_{\rm los,obs}$, but we ignore this complication here).  Here we implement this subtraction
by taking the PDF value at high velocity
$|v_{\rm sub}|= 6000-12000\ {\rm km/s}$
and subtracting this constant from the PDF.  
This subtraction is done on each \textit{individual} primary halo 
in order to reduce the
sample variance.  
Since the peculiar velocity contributions to $v_{\rm los,obs}$ are
located at much smaller values, we only consider the distribution within
$|v_{\rm los,obs}| \leq v_{\rm cut} = 2372\ {\rm km/s}$, and normalize the 
distribution within this range.  The value of $v_{\rm cut}$ corresponds to a Hubble flow difference for line-of-sight separation of 20 Mpc/h at $z=0.35$.  
The resulting distribution is shown as green (dashed) curves in
\reffig{p2d_vcut}, while its dispersion as function of $r_p$ is shown
in \reffig{hubble_sharpcut} (green points).  Clearly, the interloper
subtraction does not simply recover the distribution without Hubble
flow.  

The error bars in \reffig{hubble_sharpcut} show the uncertainty in the 
simulation measurements.  They are estimated using the 20 realizations of 
N-body simulations described in \citep{nt11_halo_halo} by subdividing each
simulation volume into 27 subvolumes (see also Fig.~1 of \cite{vlosDisp12}).  
For reference, these correspond to the expected sample variance errors for a 
survey volume of $0.056\ ({\rm Gpc}/h)^3$.

The blue points in \reffig{hubble_sharpcut} show the dispersion without
peculiar velocities, that is, only the Hubble flow contribution, and without
subtracting the constant interloper fraction.  We see that peculiar velocities
act to \emph{reduce} the velocity dispersion as compared to the pure Hubble
flow.  This is caused by the same effect responsible for the ``squashing'' of 
correlation function contours in redshift space.  Note also that
the sample variance error bars on the dispersion increase significantly 
after the subtraction of the ``interlopers''.  The question of whether this
will also be the case in the application to galaxy surveys, and whether there are more optimal
ways of dealing with interlopers, is beyond the scope of this paper.

\subsection{Incorporating Hubble flow in the model}

The phase-space distribution and line-of-sight velocity dispersion 
derived in \refsec{model} neglected the Hubble flow and assumed a fixed range 
in line-of-sight separation between primary halo and tracers.  These
effects can be straightforwardly included in the model
through \refeq{vhubble} and by extending the $z$ integration, which then leads to 
\begin{align}
p_{\rm 2D}(v_{\rm los,obs}|r_p) &\propto \int{\rm d}M_p\ n(M_p) \int {\rm d}z \int{\rm d}M_s\ 
               n(M_s) \nonumber \\*
        & \times [1+\xi_{hh}(r; M_p,M_s)]\int{\rm d}\vec{v}_{\rm halo}\ 
               p(\vec{v}_{\rm halo}) \nonumber \\*
       & \times \delta_{\rm D}(v_{\rm los,obs} - zH - \vec{v}_{\rm halo}\cdot \hat{z})\,.
\label{eqn:p2d_hubble}
\end{align}
The line-of-sight velocity dispersion (\refeq{sighalo}) then
acquires two additional contributions which scale as $z^2H^2$ and 
$2zH\langle v_{hr} \rangle\cos\phi$.  The latter term is the cross-correlation 
between the Hubble flow and mean radial infall onto the primary halo.  This
term prevents us from applying a straightforward deconvolution to reconstruct
the distribution of $v_{\rm los}$ itself.  This cross-correlation is
in fact responsible for the reduced total velocity dispersion compared to the
pure Hubble flow dispersion (red vs blue points in \reffig{hubble_sharpcut}).

Instead, we employ a ``forward-modeling'' approach
where the Hubble flow is included in the model following
\refeq{p2d_hubble}.  For this we extend the line-of-sight integration
out to the whole simulation box in the line-of-sight direction, $z_{\rm max} = 100 \Mpch$ 
(in fact the precise value of $z_{\rm max}$ does not affect the results
as long as it is sufficiently large to capture the Hubble flow contribution
for $|v_{\rm los}| \leq v_{\rm cut}$).  This results in the red line shown
in \reffig{hubble_sharpcut}, where we have used the correlation function between primary and secondary halos in real space as measured from the simulations.  We will discuss this below.

As described above, we now subtract a constant term from the $v_{\rm los}$ 
distribution, and then repeat the measurement of the dispersion.  This 
procedure is applied in the same way to the model predictions as done in 
the simulations.  The result is shown as green line in \reffig{hubble_sharpcut}.  
The agreement is very good, at approximately the same level as for the peculiar velocity dispersion when neglecting the Hubble flow contribution.  This is very important, as the goal is to use the model to infer the peculiar velocity dispersion from measurements which include the Hubble flow.

While we have employed the measured real-space correlation function of
halos here, in reality there is no direct way of accessing this correlation function.  In order to illustrate the accuracy requirements on the
correlation function, we also show the result for the Hubble-flow-subtracted 
model prediction when using the tracer correlation function predicted by the
simple halo model with exclusion employed in the model of \refsec{model} (dotted green line in 
\reffig{hubble_sharpcut}).  This correlation function departs from the 
simulation measurement by 20\% on scales $r \lesssim 8\Mpch$, with order unity
deviations for $r \lesssim 3 \Mpch$.  On the other hand, it leads to deviations of 
less than 20\% in the predicted velocity dispersion.  Thus, while a reasonably accurate
model for the correlation function (at the $\sim 5-10$\% level) is clearly necessary to incorporate the
Hubble flow properly, we expect that the desired accuracy can be met by employing
a more sophisticated model along with constraints from the observed projected 
correlation function between primary and secondary tracers.

In summary, it is clear however that the fairly simple halo model presented here captures the main properties of the phase-space distribution even when including the significant complication caused by the Hubble flow.

\section{Modified Gravity}
\label{sec:MG}

Within the halo model description of the phase-space distribution,
it is straightforward to include the main modified gravity effects that
affect the phase-space statistics.  The following ingredients of the model described in \refsec{model} will be modified:
\begin{itemize}
\item Linear matter power spectrum and growth rate:  these provide the modifications
which are seen in large-scale redshift-space distortions (RSD).  In our case, they modify the radial and transverse
velocity dispersions as well as the radial infall velocity calculated through \refeq{vspher}.  The linear power spectrum also affects the mass profile around the primary halos.
\item Virial velocity dispersion:  at a given mass, halos have a higher velocity
dispersion in modified gravity models which increase the gravitational strength.
\item Halo mass function:  a change in the mass function leads to a different mean
primary halo mass in a mass-selected (or abundance-selected) halo sample, which 
modifies the virial velocity dispersion and radial infall velocity.
\item Linear halo bias:  the halo bias quantifies the amount of surrounding mass
in our model, which is used to calculate the radial infall velocity.
\end{itemize}
The first two effects are the main modified gravity effects on velocities which we are looking for,
and provide the dominant modifications to the phase-space distribution.  However,
the effects of modified halo bias and mass function cannot be entirely
neglected.  

In the following, we will first describe the modified gravity models considered here, $f(R)$ and DGP, and then outline how the modifications to these quantities are calculated.  The choice of these models is motivated, first, by the fact
that they represent two distinct classes of modified gravity, with $f(R)$
being a special case of a chameleon model while DGP is a representative
of the Vainshtein or Galileon class.  Second, self-consistent N-body
simulations have been performed for these models, which will allow us to
quantitatively test our model.

Studying structure formation beyond linear theory in any viable modified
gravity model
is complicated by the non-linear field equations for the scalar degree of
freedom mediating the modified force.  In the models studied here, this 
non-linearity is
responsible for the chameleon and Vainshtein screening mechanisms which allow
these models to evade Solar System tests (in certain parameter regimes).  
The field equations need to be solved simultaneously with the evolution of the matter
density.   In the following we use the results of the self-consistent $N$-body simulations
of \cite{oyaizu08b} for $f(R)$ and \cite{DGPM,DGPMII} for DGP.

\subsection{$\bm{f(R)}$ gravity}
\label{sec:fR}

In the $f(R)$ model (see \cite{Nojiri:2006ri,Sotiriou:2008rp} and references therein), 
the Einstein-Hilbert action is augmented with a general function of the 
scalar curvature $R$ \cite{Caretal03,NojOdi03,Capozziello:2003tk},
 \begin{eqnarray}
S_{G}  =  \int{d^4 x \sqrt{-g} \left[ \frac{R+f(R)}{16\pi G}\right]}\,. 
\label{eqn:action}
\end{eqnarray}
Here and throughout $c=\hbar=1$.  This theory is equivalent to a 
scalar-tensor theory (if the function $f$ is nontrivial).  The additional
field given by $f_{R}\equiv df/dR$ mediates an attractive force whose
physical range is given by the Compton wavelength (inverse mass)
$\lambda_C= a^{-1}(3 d f_R/dR)^{1/2}$.  On scales smaller than $\lambda_C$,
gravitational forces are increased by a factor of 4/3, enhancing the growth of structure.   

A further important property of such models is the non-linear chameleon effect 
which shuts down the enhanced forces in regions with deep gravitational 
potential wells compared with the background field value,
$|\Psi| \gtrsim \frac32 | f_R(\bar R)|$ \cite{khoury04a,HuSaw07a}.  
This mechanism is necessary in order to pass
Solar System tests which rule out the presence of a scalar field locally.  
Thus, Solar System tests conservatively constrain the amplitude of the background field
to be less than typical cosmological potential wells today ($\sim 10^{-6} - 10^{-5}$).  

In this paper, we will choose the functional form introduced by Hu \& Sawicki
\cite{HuSaw07a}:
\begin{equation}
f(R) = - 2\Lambda \frac{R}{R+\mu^2},
\end{equation}
with two free parameters, $\Lambda$, $\mu^2$.  Note that as $R\rightarrow 0$,
$f(R)\rightarrow 0$, and hence this model does not contain a cosmological
constant.  
Nevertheless, as $R \gg \mu^2$,  the function $f(R)$ can be approximated as
\begin{equation}
f(R) = -2 \Lambda - f_{R0} \frac{\bar R_0}{ R} \,,
\label{eqn:fRapprox}
\end{equation}
with $f_{R0}= -2 \Lambda \mu^2/\bar R_0^2$ replacing $\mu$ as the second 
parameter of the model.  Here we define $\bar R_{0}=\bar R(z=0)$, 
so that $f_{R0}= f_{R}(\bar R_{0})$, where overbars denote the quantities of 
the background spacetime.  Note that $f_{R0} < 0$ implies $f_R < 0$ always, as 
required for stable cosmological evolution.  If $|f_{R0}| \ll 1$, 
the curvature scales set by $\Lambda ={\cal O}(R_0)$
and $\mu^2$ differ widely and hence the $R \gg \mu^2$ approximation is valid 
today and for all times in the past.

The background expansion history thus mimics $\Lambda$CDM with $\Lambda$ as
a true cosmological constant to order $f_{R0}$.   Therefore in the limit 
$|f_{R0}| \ll 10^{-2}$, the $f(R)$ model and $\Lambda$CDM are essentially 
indistinguishable with geometric tests.  Gravitational forces are unmodified
on scales larger than $\lambda_C$, while they are enhanced by a factor of
4/3 on scales below $\lambda_C$ (in regions where the chameleon mechanism
is not active).  For reference, for the model adopted here, $\lambda_C(z=0)\simeq 23 \Mpch (|f_{R0}|/10^{-4})^{1/2}$.   Note that $\lambda_C$ is a function of 
redshift.  Correspondingly, the linear growth rate is strongly
scale-dependent on small scales, while it matches $\Lambda$CDM on scales
larger than $\lambda_C$ \cite{HuSaw07a}.  

While we choose a specific functional form for $f(R)$ here, it
is straightforward to map constraints onto different functional forms
(see \cite{FerraroEtal} for details).  In the following, for notational simplicity
$f_{R0}$ will always refer to the absolute value of the field amplitude today.  

\subsection{DGP}
\label{sec:DGP}

In the DGP braneworld scenario \cite{DGP1}, matter and radiation live
on a four-dimensional brane in five-dimensional Minkowski space.  The action
is constructed so that on scales larger than the crossover scale
$r_c$, gravity is five-dimensional, while it becomes four-dimensional
on scales smaller than $r_c$.  This model admits a homogeneous cosmological
solution on the brane which obeys a modified Friedmann equation \cite{Deffayet01}:
\be
H^2 \pm \frac{H}{r_c} = 8\pi G\: [\bar\rho_m + \rho_{\rm DE}].
\ee
The sign on the l.h.s. is determined by the choice of embedding of the brane.  
The negative sign is called the \textit{self-accelerating} branch, since
it allows for accelerated expansion even in the absence
of a cosmological constant.  The positive sign is called the \textit{normal}
branch, which does not exhibit self-acceleration.  Here, we consider
models of both branches (see \cite{DGPM,DGPMII,DGPhalopaper}):  
a self-accelerating model without a $\Lambda$ term ($\rho_{\rm DE}=0$), {\it sDGP}, 
where $r_c\sim 6000$~Mpc is
adjusted to best match CMB and expansion history constraints \cite{FangEtal};  
and normal-branch models with a dark energy component $\rho_{\rm DE}$ adjusted 
so that the expansion history is exactly $\Lambda$CDM \cite{DGPMII}.  In that
case, $r_c$ is a free parameter, and we chose values of 500~Mpc ({\it nDGP--1})
and 3000~Mpc ({\it nDGP--2}).  We emphasize that while the DGP model
itself is highly constrained (if one does not combine it with a tailored
Dark Energy component), many recent developments in the context of 
massive gravity and degravitation (e.g., \cite{cascade,degrav,galileon,dRG,dRGT}) are expected to behave phenomenologically 
similar to DGP in the large-scale structure regime; in particular, all these
models share the Vainshtein screening mechanism.

On sub-horizon scales, and scales smaller than the crossover scale $r_c$,
DGP braneworld models can be accurately described
as a scalar-tensor theory \cite{NicRat}, where the brane-bending mode $\varphi$ mediates
an additional attractive (normal branch) or repulsive (self-accelerating branch)
force.  Gravitational forces in DGP are governed by:
\be
\bm{\nabla}\Psi = \bm{\nabla}\Psi_N + \frac{1}{2}\bm{\nabla}\varphi.
\label{eq:PsiDGP}
\ee
The $\varphi$ field is sourced by matter overdensities similarly to the usual
GR potentials, but has quadratic self-interactions which suppress
the field once density contrasts become non-linear (e.g. \cite{KoyamaSilva}).  
In the linear regime, gravitational forces are modified by a scale-independent factor $1+(3\beta)^{-1}$, where $\beta$ is a redshift-dependent function of order $H r_c$ ($- H r_c$ for the self-accelerating branch).  

When the density field becomes non-linear, the derivative self-interactions of the $\varphi$ field become important.  Analytical solutions to the full $\varphi$ equation do not exist in general, however the case of a spherically symmetric
mass is solvable in terms of closed expressions \cite{LueEtal04,KoyamaSilva}.  In particular, one finds that the modified force is suppressed within the Vainshtein radius $r_*$ given by
\be
r_* = \left(\frac{16 G  M r_c^2}{9\beta^2}\right)^{1/3}\,.
\ee
On small scales $r \ll r_*$, modified forces are suppressed by 
$(\varepsilon\bar\d)^{-1/2}$, where $\bar\d=\d\rho(<r)/\bar\rho_m$ is the average overdensity within $r$ and $\varepsilon$ is a parameter of order unity for the models considered here \cite{dynamicalmass}.

\subsection{Simulation measurements}
\label{sec:MGsim}

The simulations for $f(R)$ gravity and the halo finding applied are described
in \cite{oyaizu08b,halopaper}.  For DGP, they are described in \cite{DGPM,DGPMII}.  For both models, we use 6 realizations (3 in case of the nDGP models) of a $L=256\Mpch$ simulation box run on a fixed $512^3$ grid.  The primary halos are identified using a spherical overdensity algorithm, and have masses $M_{300m} > 10^{14}h^{-1} M_\odot$ ($M_{200m} > 10^{14} h^{-1} M_\odot$ in case of DGP).  Due to the limited simulation volume and resolution, we only consider the dark matter phase space around these halos.  For each model and realization, we measure the phase space as described in \refsec{model}, and determined the RMS peculiar velocity dispersion $\s_{v_{\rm los}}(r_p)$.  We then take the ratio between the modified gravity result and the GR result, and average these ratios over the six (three) realizations.  The points in \reffig{sigvlosfR} (\reffig{sigvlosDGP}) show
the results for $f(R)$ (DGP), respectively.  The error bars are bootstrap 
errors on the mean ratio from the realizations.

\subsection{Halo model}
\label{sec:MGmodel}

\subsubsection{Linear growth rate and matter power spectrum}

We solve the linear growth factor for both $f(R)$ and DGP models by integrating
the modified linear perturbation equations.  In case of $f(R)$, the growth
factor is scale-dependent, and we use the small-scale limit (corresponding to
modified forces throughout) in \refeq{vspher}.  This approximation will overestimate the $f(R)$ modifications to the mean radial infall somewhat.  We do use the full scale-dependent growth factor for all other quantities.  Further,
the modified spherical collapse threshold (see below) is used in 
\refeq{vspher}.

\subsubsection{Cluster abundance and clustering}

The abundance of dark matter halos (mass function) and
their clustering (halo bias) in the $f(R)$ simulations was studied
in \cite{halopaper}, and correspondingly for DGP in \cite{DGPhalopaper}.    
Analytical approximations using the excursion set formalism have been studied \cite{le2012,ll2012a,ll2012b,lucas2013}.
We use a simple model developed in these papers based on spherical 
collapse and the peak-background split in order to predict the 
cluster abundance and their linear bias.

We employ the Sheth-Tormen prescription
for the comoving number density of halos per logarithmic interval in the 
\emph{virial} mass $M_{\rm v}$, given by
\begin{align}
n_{\rm v}^{\rm (ST)} \equiv
\frac{d n}{d\ln M_{\rm v}} &= {\bar \rho_{\rm m} \over M_{\rm v}} f(\nu) {d\nu \over d\ln M_{\rm v}}\,, 
         \label{eqn:massfn}
\end{align}
where the peak threshold $\nu = \delta_c/\sigma(M_{\rm v})$ and 
\begin{eqnarray}
\nu f(\nu) = A\sqrt{{2 \over \pi} q\nu^2 } [1+(q\nu^2)^{-p}] \exp[-q\nu^2/2]\,.
\end{eqnarray}
Here $\sigma(M)$ is the variance of the linear density field 
convolved with a top hat of radius $r$
that encloses $M=4\pi r^3 \bar \rho_{\rm m}/3$ at the background density
\begin{eqnarray}
\sigma^2(r) = \int \frac{d^3k}{(2\pi)^3} |\tilde{W}(kr)|^2 P_L(k)\,,
\label{eqn:sigmaR}
\end{eqnarray}
where $P_L(k)$ is the linear power spectrum (either
in $\Lambda$CDM or in modified gravity) and $\tilde W$ is the Fourier transform
of the top hat window.  The normalization constant $A$ is chosen 
such that $\int d\nu f(\nu)=1$, and we adopt  $p=0.3$, $q=0.75$, and
$\delta_c=1.673$ where the latter is obtained from a numerical spherical
collapse calculation \cite{halopaper}.  
The virial mass is defined as the mass enclosed at 
the virial radius $r_{\rm v}$, at which the average density is $\Delta_{\rm v}$
times the mean density.   $\D_{\rm v}$ is obtained from the spherical collapse calculation described in \cite{halopaper}.  We transform the virial mass to the 
desired overdensity criterion $\Delta$ assuming a 
Navarro-Frenk-White 
\cite{NavFreWhi97} density profile \cite{HuKravtsov}, using the
mass-concentration relation of \cite{Buletal01} (see also \refapp{vm}).  We thus obtain the mass function of halos
in the ST prescription, $n^{\rm (ST)}$, from $n_{\rm v}^{\rm (ST)}$.  

The effects of modified gravity enter in two ways in this prescription:  
first, we use the linear power spectrum for the specific model in 
\refeq{sigmaR}.  Second, we assume modified spherical collapse parameters $\d_c,\,\D_{\rm v}$.  
In the case of $f(R)$, we consider two limiting cases of spherical collapse: the first does not involve any force modification (collapse parameters as in $\Lambda$CDM), corresponding to the case where the collapsing region is always larger than the Compton wavelength $\lambda_C$ of the field.  In the second case we rescale the gravitational 
constant by 4/3 during the collapse calculation as well as the corresponding 
linear growth extrapolation to obtain $\d_c$.    
This corresponds to the case where the collapsing region is always
smaller than $\lambda_C$.  
For DGP on the other hand, we use the exact solution of the brane-bending
mode equation for a spherical tophat to evaluate the modified force during
collapse \cite{DGPhalopaper}, and use the spherical collapse parameters obtained numerically.  Note that we only use the ST prescription and spherical collapse to predict 
the relative \emph{enhancement} of the halo abundance in modified gravity.

In addition to the halo abundance, modified gravity also affects 
the clustering of halos.  This effect comes from two sources:  first,
the matter power spectrum is modified.  
Second, the linear bias $b_1(M,z)$ of halos at a given mass
$M$ and redshift $z$ is modified.  For example, in $f(R)$ and normal-branch
DGP, halo bias is reduced at fixed mass for massive halos since these are 
less rare than in GR due to the increased mass function.  In order to model
the modified gravity effects on halo bias, we use the peak-background split
bias derived from the Sheth-Tormen mass function [\refeq{massfn}],
\be
b_L(M) = 1 + {q \nu^2 -1 \over \delta_c}
        + { 2 p \over \delta_c [ 1 + (q \nu^2)^p]}\,,
\label{eqn:biasST}
\ee
where $\nu, q, p$ are defined after \refeq{massfn}.  Note that $\nu$ is given 
in terms of the virial mass $M_{\rm v}$, and thus for a given mass and
redshift $\nu$
differs in modified gravity due to both the modified spherical collapse parameters
and the different linear power spectrum.  

\subsubsection{Virial velocities}
\label{sec:MGvirial}

In modified gravity, the dynamics of non-relativistic bodies (such as
galaxies or dark matter) within the potential well of a body of mass
$M$ is modified from that of GR due to the presence of the additional 
scalar degree of freedom.  As shown in \cite{dynamicalmass}, the
mean velocity dispersion for a halo of fixed mass predicted by the virial theorem 
in modified gravity is related to that in GR through
\be
\sigma_{\rm vir, MG}^2 = \bar g \: \sigma_{\rm vir, GR}^2,
\label{eqn:Mdyn}
\ee
where $\bar g$ is a weighted integral of the force modification
over the object which describes the effect on the virial equation.  
Assuming an NFW profile \cite{NavFreWhi97} for the primary halos, 
we have 
\be
\bar g = \frac{\int_0^{r_{\rm v}} dr\:r^2\:\rho_{\rm NFW}(r)\:g(r)\:  r\, d\Psi_N/dr}
{\int_0^{r_{\rm v}} dr\:r^2\:\rho_{\rm NFW}(r)\: r\, d\Psi_N/dr},
\label{eqn:gbar}
\ee
where $\Psi_N$ is the Newtonian potential of the halo, found by
solving (see \cite{dynamicalmass} for an explicit expression)
\be
\nabla^2\Psi_N = 4\pi G\rho_{\rm NFW},
\ee
and $g(r)$ is the force modification. In order to calculate the force
modification, we have to solve the field equations for the scalar degree
of freedom for an NFW halo \cite{dynamicalmass}.  In case of $f(R)$, this
can only be done in a computationally expensive numerical calculation.  
Hence we instead use a simple model which describes the exact results 
reasonably well \cite{dynamicalmass}.  
Specifically, we assume that the chameleon field is only sourced by
the ``thin shell'' of mass not screened by the chameleon effect [see Eq.~(40) in
\cite{dynamicalmass}].  $\bar g$ then interpolates between $4/3$ for
low-mass halos where the chameleon effect is not active, to 1 for halos
which are completely screened.  In case of DGP, it is possible to
solve the equation for the brane-bending mode for a spherically
symmetric halo analytically [Eq.~(47) in \cite{dynamicalmass}], and we use
this result in \refeq{gbar}.   

\begin{figure}
\centering
\includegraphics[width =0.48\textwidth, clip=true, trim=0.8cm 1.3cm 1.8cm 0.8cm]{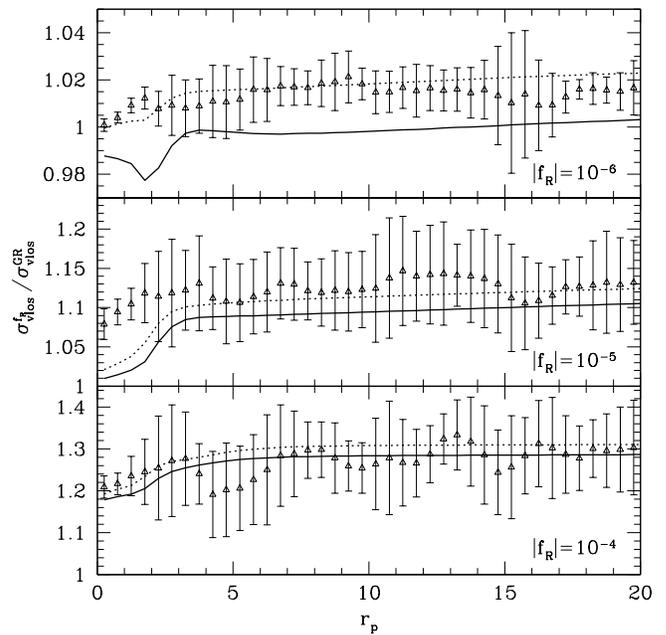}
\caption{Ratio of the RMS line-of-sight velocity dispersion $\s_{v_{\rm
 los}}$ in $f(R)$ with respect to $\Lambda$CDM.  The panels show
 different values for the background field amplitude today.  Open
 symbols show the results of the full simulations.  The solid lines show
 the model prediction using modified spherical collapse parameters,
 while the dotted line denotes the prediction using the 
unmodified spherical collapse
parameters.
}
\label{fig:sigvlosfR}
\end{figure}

\begin{figure}
\centering
\includegraphics[width =0.48\textwidth, clip=true, trim=0.8cm 1.3cm 1.8cm 0.8cm]{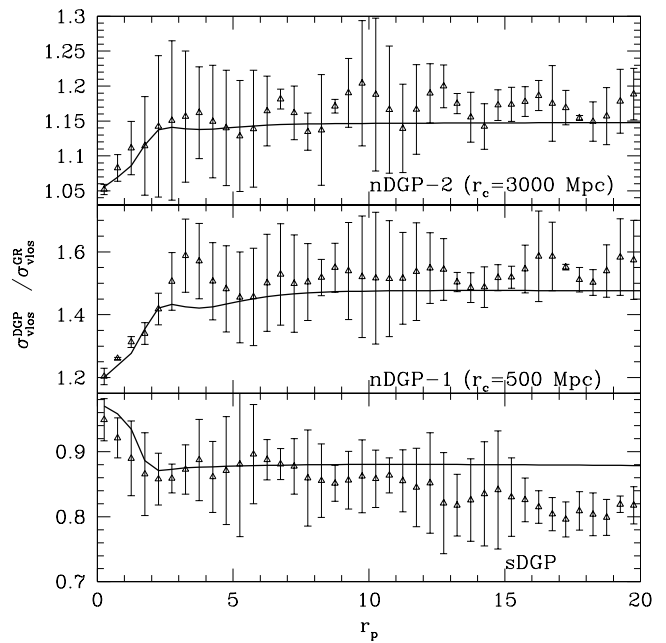}
\caption{Same as \reffig{sigvlosfR}, but for DGP models.  The top two panels show normal-branch models, while the lower panel shows the self-accelerating model where gravity and velocities are suppressed with respect to GR.  The solid line shows the model prediction.
}
\label{fig:sigvlosDGP}
\end{figure}

\subsection{Results}

\reffig{sigvlosfR} shows the ratio of the velocity dispersion $\s_{v_{\rm los}}$
in $f(R)$ compared to $\Lambda$CDM as function of $r_p$.  \reffig{sigvlosDGP}
shows the same for DGP (more precisely, here the modification is with respect
to GR models with the same expansion history).  The model predictions are
shown as lines.

In case of $f(R)$, we see that the two spherical collapse cases roughly 
bracket the simulation results, where the unmodified collapse parameters
(dotted lines) are closer to the simulation results for $|f_{R0}| \leq 10^{-5}$.  
Some discrepancy can be seen at $r_p < 3\Mpch$ in case of the intermediate field value.  Large discrepancies were found around the same scale in the GR case (\refsec{model}).  Another possible explanation is that the approximate treatment of chameleon screening of the primary halo (\refsec{MGvirial}) overestimates the screening effect, as already found in \cite{dynamicalmass}.  The weak-field case $|f_{R0}| = 10^{-6}$, in which the modified gravity effects are strongly suppressed further, exemplifies the chameleon mechanism.  In this case, the primary halos are fully screened resulting in essentially unmodified
infall velocities.  The velocity dispersion of dark matter particles surrounding
the primary halos is however still modified at the few percent level.  
For the smaller field values $|f_{R0}| \leq 10^{-5}$, the prediction using the unmodified spherical
collapse parameters gives a somewhat better fit to the simulation results.  
The spherical collapse parameters mostly affect the mass function (except for
the smallest field case, where the 1\% reduction of the radial infall motion 
when using the modified $\delta_c$ also becomes noticeable). 
As found in \cite{halopaper}, the unmodified spherical collapse parameters
give a somewhat better fit to the halo mass function in $f(R)$ for $|f_{R0}| \leq 10^{-5}$, which provides a likely explanation for the results seen in \reffig{sigvlosfR}.

For DGP (\reffig{sigvlosDGP}), the model captures the simulation results well on all scales, including the transition region to the 1-halo regime.  Significant $15-20$\% effects on the velocity dispersion $\s_{v_{\rm los}}$ are seen even for horizon-scale cross-over scales $r_c \gtrsim 3000$~Mpc;  note that this corresponds to $30-40$\% effects on the variance.  There is an unexpected discrepancy on large scales in the sDGP model which could be a statistical fluctuation (note that the errors are correlated and computed from only 6 realizations).

In summary, the halo model appears to capture the effects of modified
gravity on the line-of-sight velocity dispersion both qualitatively and
quantitatively for a range of models and parameters.  We thus expect
that this model can be exploited to derive constraints on modified gravity
from observations of the phase-space distribution in galaxy surveys.

\section{Discussion}
\label{sec:discussion}

The velocities of large-scale structure tracers in principle provide 
rich information on the growth of structure by circumventing the issue
of bias which affects the relation between tracer and matter densities.  
This dynamical information is of particular interest in models with modified
gravity since the relation of dynamical mass and the gravitational lensing 
mass is generically modified in theories beyond General Relativity.  

On large scales above $\sim 30\Mpch$, the velocities are best probed through
redshift-space distortions of correlation functions and power spectra.  
On small scales below $\sim 2 \Mpch$, virial velocities within massive halos
(galaxy clusters) can be probed by a wide range of observables such as
galaxy velocity dispersion, X-ray emission, and the Sunyaev-Zel'dovich effect.  
However, these small scales are also affected by baryonic effects, 
tidal friction, and non-thermal pressure support.

Here, following previous work \cite{vlosDisp12}, we have studied the
phase-space distribution around massive halos as probe of 
velocities in the intermediate range of scales $\sim 2-15\Mpch$, partially
bridging the gap between redshift-space distortions and virial velocities.  
We describe an analytic model to compute 
the phase-space distribution of different tracers around massive clusters.  
This model is based on the halo model where the tracer velocities are composed
of a 1-halo (virial motion within halos) and 2-halo contribution (relative
center-of-mass motion of halos).  Specifically, we use dark matter particles 
as well as intermediate-mass halos as the tracers, although the formalism
can be straightforwardly generalized to model galaxy motions.

For the phase-space distribution of dark matter particles around primary halos,
the line-of-sight velocity has contribution from both 
virial velocity and halo-halo pairwise velocity.  This will in general
also be the case for galaxies.  The phase-space distribution of secondary
halos on the other hand only receives contributions from the halo-halo
pairwise velocity.  The basic ingredients that enter our model
are the virial velocity dispersion \cite{evrardetal08} for the 1-halo term;
the mean radial infall velocity; the Gaussian dispersion of the radial and
tangential velocities; and the mass function and bias of halos as function
of mass.  For the mean radial infall, we adopt an empirical prescription 
which is motivated by the spherical collapse model.  We apply the 
characteristic function approach to efficiently compute the line-of-sight 
velocity dispersion.

The resulting phase-space distribution and the velocity 
dispersion match the measurement from $N$-body simulations very well on
scales above $\sim 2.5\Mpch$, for both dark matter and secondary halos.  
On smaller scales, the radial infall
velocity is overpredicted resulting in a worsening match to simulation
results.  We thus expect that the most straightforward way to significantly 
improve the model is to develop a more realistic prescription for the 
radial infall motion.  

We also discuss the effects of the Hubble flow, i.e. the fact that in reality 
we can only measure 
redshift differences which receive contributions from both 
peculiar velocity and differential Hubble flow.  The Hubble flow contribution 
dominates the peculiar velocity when the line-of-sight separation 
of the halo pair exceeds several $h^{-1}$Mpc.  
We show that our model can take these effects into account, and still matches 
the simulation measurements very well if the real-space correlation function
between primary halos and tracers is accurate at the level of $5-10$\%.  

During the preparation of this work, another independent study by 
\citet{zuweinberg2012} also investigates the modeling of the infall region
on massive halos.  Our approach agrees with theirs in separating the velocity 
contribution into virial velocity and infall velocity.  However, the
two studies are complementary both in methods and in goals.  While \citet{zuweinberg2012} aim to reconstruct the infall velocity in the context of the
standard $\Lambda$CDM model, our goal is to use the phase-space distribution
as a probe of modified gravity models.  Further, the model of \cite{zuweinberg2012} involves  7 fitting parameters for each primary and tracer sample.  
On the other hand, our approach only involves two matching scales for the
entire distribution: the scales at which the infall velocity and tangential
dispersion are matched to the linear theory predictions.  Moreover,
we have chosen both to be equal and identical for all tracer samples.

While our model predictions successfully match the measurement well, 
there are numerous improvements that can be done. 
Our model description on the halo-halo pairwise velocity distribution, 
while based upon the spherical collapse model and the linear theory 
prediction, is empirical. The discussion in the appendix shows 
that while this empirical approach can describe the peaks of both
the radial and the tangential component distributions, 
the agreement in the high velocity tail is not as good. 
Furthermore the empirical model assumes the two distributions are 
independent Gaussian distributed 
(for the tangential component it is its projection 
onto the line-of-sight direction), which is only a crude approximation.
While this approximation works well for the phase-space distribution 
of halo-dark matter pairs and the line-of-sight velocity dispersion of
the halo-halo pairs, the discrepancies do matter when 
higher order moments are considered.

Another improvement under investigation is including the 
virial velocity of galaxies in the calculation.  
These contributions can be minimized by selecting LRGs or BCGs which 
reside close to the secondary halo center.  Alternatively, they can be
modeled by incorporating the motions of galaxies within their halos
in a halo occupation distribution approach.

\acknowledgments

We thank Anna Cabr\'e and Bhuvnesh Jain for helpful discussions.  
F.~S. is supported by NASA through Einstein Postdoctoral Fellowship
grant number PF2-130100 awarded by the Chandra X-ray Center, which is
operated by the Smithsonian Astrophysical Observatory for NASA under
contract NAS8-03060.
T.~N. is supported by Japan Society for the Promotion of Science (JSPS).
M.~T. is supported in part by the FIRST
program ``Subaru Measurements of Images and Redshifts (SuMIRe)'',
CSTP, Japan, World Premier International Research Center Initiative
(WPI Initiative), MEXT, Japan, 
 and by Grant-in-Aid for Scientific Research from the JSPS
Promotion of Science (No.~23340061).
\bibliography{Astro}

\appendix
\section{Virial motion}\label{sec:vm}
Virial motion inside dark matter halos has been studied in
\citep{bn98,evrardetal08,cuestaetal08}. 
In this work we will use the fitting formula
provided in \citet{evrardetal08} who suggested that the virial
motion can be approximated by the Maxwellian distribution (which is true
for isothermal sphere) and the dispersion is related to the virial
mass by 
\begin{equation}
\sigma_{\rm DM}(M,z) = \sigma_{\rm DM,
  15}\left[\frac{(1+z)^{3/2}M_{\rm 200b}}{10^{15}M_{\rm
      sun}}\right]^{\alpha},
\label{eqn:sigM}
\end{equation}
where $M_{\rm 200b}$ is the mass of the halo enclosed in a radius containing
a mean density of $200 \bar\rho_m$ and $z$ is the redshift. 
They found that $\sigma_{\rm DM,15} = 880\ {\rm km/s}$ and $\alpha = 0.355$
best fit the measurements
from a series of $N$-body simulations.
The projection of the virial velocity along the line-of-sight
direction is hence a Gaussian distribution with the variance given by
\refeq{sigM}.
In this work we define halo to have average density that is 178 times the
background density $\bar{\rho}_m$. This halo mass ($m_{178b}$) definition is related to halo whose
density is $\Delta$ times a density $\varrho$
through
\begin{equation}
m = m_{178b} \frac{f(x)}{f(c)},
\end{equation}
where $x$ is the non-zero solution of 
\begin{equation}
\frac{\Delta \varrho x^3}{f(x)} = \frac{178 \bar{\rho}_mc^3}{f(c)},
\end{equation}
and 
\begin{equation}
f(y) = \ln(1+y) -\frac{y}{1+y}.
\end{equation}
The above fitting formulae have $(\Delta,\varrho) = (\Delta_{\rm
  vir},\rho_{\rm crit})$ and $(200,\bar{\rho}_m)$ respectively.
The concentration parameter $c(m,z)$ is obtained by the fitting formula
\cite{Buletal01}
\begin{equation}
c(m,z) = \frac{9}{1+z}\left[\frac{m}{m_*(z)}\right]^{-0.13},
\end{equation}
where $m_*(z)$ is the characteristic mass scale at $z$.


\section{Halo-halo pairwise velocity distribution}\label{sec:hhpvd}
There are two contributions (for dark matter as tracers) 
for the 2-halo term. We assume the virial motion within the 
secondary halo is also described by the fitting formula in the previous 
section -- it only depends on the mass of the secondary halo.
We describe an approximation that leads to an analytic expression 
for the mass-weighted virial velocity in secondary halos in \refapp{Qvir}.  
In this section we describe our model for the halo-halo pairwise velocity.  

The spherical collapse model relates the linear overdensity
$\delta_l$ and the evolved overdensity $\delta_{\rm NL}$ 
and we will use the following approximation
\citep{b94,rks98}:
\begin{equation}
\frac{M(< r)}{\bar{\rho}_m V} \equiv 1 + \delta_{\rm NL} = \left( 1- \frac{D(t)\delta_l}{\delta_c}\right)^{-\delta_c},
\label{eqn:SC}
\end{equation}
where $\delta_c = 1.686$ in EdS universe and its value is weakly
dependent on cosmology. 
The above expression allows one to estimate the radial infall velocity as the
time derivative of the radius of the mass shell, $v_{\rm SC}(r) = \dot r$
(see, for example, Eq.~(6) in \citep{lamshethred} with $\lambda_k = \delta_l/3$),
\begin{equation}
v_{\rm SC}(r) = -\frac{H(z)}{1+z} r \frac{f(z)}{3}
\delta_c \left[(1+\delta_{\rm NL})^{1/\delta_c} - 1\right],
\label{eqn:vspher}
\end{equation}
where the subscript SC stands for spherical collapse, $H(z)$ is the Hubble parameter at redshift $z$,
$f(z) = {\rm d}\ln D/{\rm d}\ln a$  is the rate of change of the
linear growth function. $\delta_{\rm NL}$ is the nonlinear density
contrast that includes all the mass within the radius $r$.  
While \refeq{SC} is a fitting formula, we have checked that \refeq{vspher} agrees with the numerical calculation within a few percent for $\Lambda$CDM.  
Overdense regions contract (negative radial velocity)
while underdense regions expand (positive radial velocity).
In this study we approximate the mass shell surrounding the primary halo 
using the halo-matter correlation function.   The total enclosed
mass is then given by the sum of the primary halo mass and the mass shell:
\begin{equation}
M(<r) = M_{\rm halo} + M_{\rm shell}(<r), 
\end{equation}
where
\begin{align}
M_{\rm shell}(<r)\equiv\:& \Theta(r-r_{\rm vir}) \vs
&\hspace{-2em}\times \int^{r}_{r_{\rm vir}} 4\pi r^{\prime 2}dr' \ [1 +
\xi_{h\delta}(r';M_{\rm halo})]\bar{\rho}_m,
\label{eqn:mshell}
\end{align}
where $\Theta(x)$ is the Heaviside step function,
and we approximate the halo-matter correlation function 
 given by \refeq{xihm}.

\begin{figure}
\centering
\includegraphics[width=0.45\textwidth]{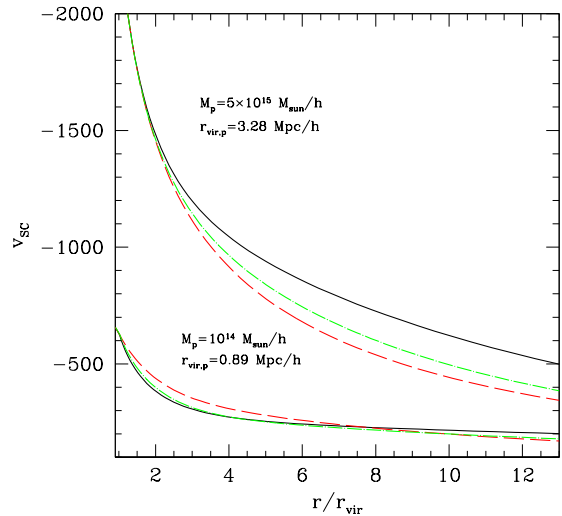}
\caption{Velocity infall as predicted by the spherical collapse
  model (\refeq{vspher}). 
The upper set of curves is the predictions for $m=5\times
  10^{15} M_{\odot}/h$ while the lower set is for $m=10^{14} M_{\odot}/h$.
  Different curves in each set show infall velocity using
  different matter correlated functions: solid (linear correlation
  function), dashed \citep[halo model, see ][]{haloreview}, 
  dot-dashed \citep[halofit, see ][]{smithetal}.}
\label{fig:vinfall}
\end{figure}
\reffig{vinfall} shows the spherical collapse model predicted
radial velocity as a function of scale for two different primary
masses. Three different matter-matter correlation functions are used
for comparison. Note that the spherical collapse model does not
predict any tangential velocity or scatter in the radial velocity
which depends deterministically  on $M(<r)$.

Before comparing the halo pairwise radial velocity measured from the
numerical simulations to the spherical collapse model prediction, we
now first briefly describe the linear theory prediction for the
pairwise
halo-halo velocity distribution. \citet{sz09} studied the velocity
correlation function and found that including the pair weighting is
important. More sophisticated empirical modelings are done in
\citet{hkysj05,tinker07,rw11}.
The linear halo-halo pairwise velocity is given by modifying equation~(17) in
\citet{lnyfnlvel}, which describes the linear pairwise velocity 
for dark matter particle pairs.
For a halo-pair of mass $M_p$ and $M_s$ separated by $r$,
\begin{widetext}
\begin{equation}
p_{hh}(\vec{v}_{\rm halo}|r, M_p,M_s)  = 
  \frac{1 + \xi_{hh}(r;M_p,M_s) + H_1(\nu_{hr})\beta_{100} + H_2(\nu_{hr})\beta_{200}}{1+ \xi_{hh}(r;M_p,M_s)}p_0(\vec{v}_{\rm halo}), 
\label{eqn:linpairwise}
\end{equation}
\end{widetext}
Where $p_0(\vec{u})$ is 
the linear velocity difference distribution.  It is given by a product 
of a Gaussian distribution (with zero mean) corresponding to 
the radial component and a Rayleigh distribution corresponding to 
the quadrature sum of the two tangential components 
(due to symmetry we are free to choose one of the tangential components to be 
perpendicular to the line-of-sight direction and the other tangential component is
a Gaussian distribution with zero mean as discussed in \refsec{model}).  
Further, $H_1$ and $H_2$ are Hermite polynomials and
$\nu_{hr} = v_{hr}/\sigma_{u_{hr}}$.
The variances of the linear velocity difference distribution and other parameters
in \refeq{linpairwise} are
\begin{widetext}
\begin{align}
\sigma_{u_{hr}}^2 & = \frac{1}{3\pi^2}\dot{D}^2 \int{\rm d}k\ 
     P_{\delta\delta} (k) \left[\frac{W(k,M_p)^2+W(k,M_s)^2}{2} 
              -3W(k,M_p)W(k,M_s)j_0(kr) 
                  + 6W(k,M_p)W(k,M_s)\frac{j_1(kr)}{kr}\right], \nonumber \\
\sigma_{u_{ht}}^2 & = \frac{1}{3\pi^2}\dot{D}^2 \int{\rm d}k\ 
     P_{\delta\delta} (k) \left[\frac{W(k,M_p)^2+W(k,M_s)^2}{2} 
                - 3W(k,M_p)W(k,M_s)\frac{j_1(kr)}{kr}\right], 
\label{eqn:siguht} \\
\beta_{100} &= \frac{\langle u_{hr}\delta_h(M_p)\rangle + \langle u_{hr}\delta_h(M_s)\rangle}{\sigma_{u_{hr}}}, \qquad 
    \beta_{200} = \frac{\langle u_{hr}\delta_h(M_p)\rangle\langle u_{hr}\delta_h(M_s)\rangle}{\sigma_{u_{hr}}^2}, \nonumber \\
& \langle u_{hr}\delta_h(M_p)\rangle = \langle  u_{hr}\delta_h(M_s) \rangle =
  -\frac{1}{2\pi^2}\dot{D}\int {\rm d}k\ 
        P_{\delta\delta}(k)kj_1(kr)W(k,M_p)W(k,M_s), \nonumber
\label{eqn:linvdiff}
\end{align}
\end{widetext}
where $\dot D = dD/dt = H D f$ is the time derivative of the linear growth factor.
We compute the means and dispersions of the radial
velocity 
$v_{hr}$
and the tangential velocity $v_{ht}$
from the pair-weighted distribution 
with a weighting factor of $n(M_p)n(M_s)[1 +
\xi_{hh}(r,M_p,M_s)]$ to take into account the halo number
density and clustering. 

\begin{figure}
\centering
\includegraphics[width=0.5\textwidth]{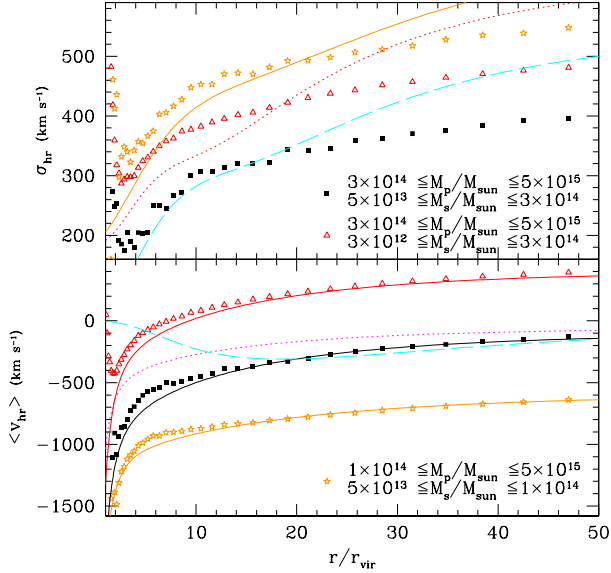}
\caption{Halo-halo pairwise velocity (radial component): 
  top and bottom panels show the dispersion and mean, respectively of the 
  radial component of the halo-halo pairwise velocity
  measured with different mass ranges (see legends for description).
  The curves in the upper panel show the linear theory 
  prediction on the dispersion (the red and the orange sets 
  are displaced by 80 km/s and 160 km/s respectively). 
  In the lower panel, the cyan dashed and the magenta dotted curves 
  are the predictions of the linear theory and the spherical collapse
  model, respectively, for the halo mass range represented by the solid black squares.   
  The solid black, red, and orange lines show the rescaled spherical collapse
  prediction (see text; the red and orange sets are displaced by $\pm 500\ {\rm km/s}$ for clarity). 
}
\label{fig:vhr}
\end{figure}

\reffig{vhr} shows the mean $\< v_{hr} \>$ and dispersion $\sigma_{hr}$ 
of the radial halo-halo pairwise velocity. 
We identified halo pairs in the numerical simulations 
with different mass ranges for primary and secondary halos (see legends
for description).
The bottom panel shows the mean radial velocity (the red and 
the orange sets are displaced $\pm 500\ {\rm km/s}$ respectively).
The cyan dash curve is the linear theory prediction for 
halo mass ranges represented by the solid black squares while
the magenta dotted curve is the prediction from spherical collapse model 
(where the halofit correlation function is used).

The linear theory predictions match the mean radial infall relatively 
well for large separation, but the predictions are wrong at
small separation. On the other hand the prediction from the spherical 
collapse model has the correct scale dependence but the magnitude is 
not matching the measurement. 
Several reasons can contribute to this, although likely the most important
factor is the assumption of zero angular momentum in the spherical
collapse model.

We construct an empirical model to describe the mean radial infall 
of halo pairs: we take the predictions from the spherical collapse model
and rescale them by a \emph{constant} factor derived by the linear theory 
prediction at large separation (we take $r/r_{\rm vir}=20)$.
The solid curves are our empirical model predictions. They 
match the measurements for separation larger than 10 virial radii
of the primary halos but overpredict $| \< v_{hr} \> |$ for smaller separations.

The upper panel of \reffig{vhr} shows the dispersion of the
radial infall. Symbols are measurements from numerical 
simulations with different halo mass ranges and the curves
are the linear theory predictions. 
Linear theory can only qualitatively match the measurements and there is a
10-20\% discrepancies between them (at intermediate separation).
Nonetheless we will use the linear theory prediction for the
radial infall dispersion.  This is justified by its relatively 
small contribution to the line-of-sight velocity dispersion 
(see \reffig{sigvlos_GR_split}).

\begin{figure}
\centering
\includegraphics[width=0.4\textwidth]{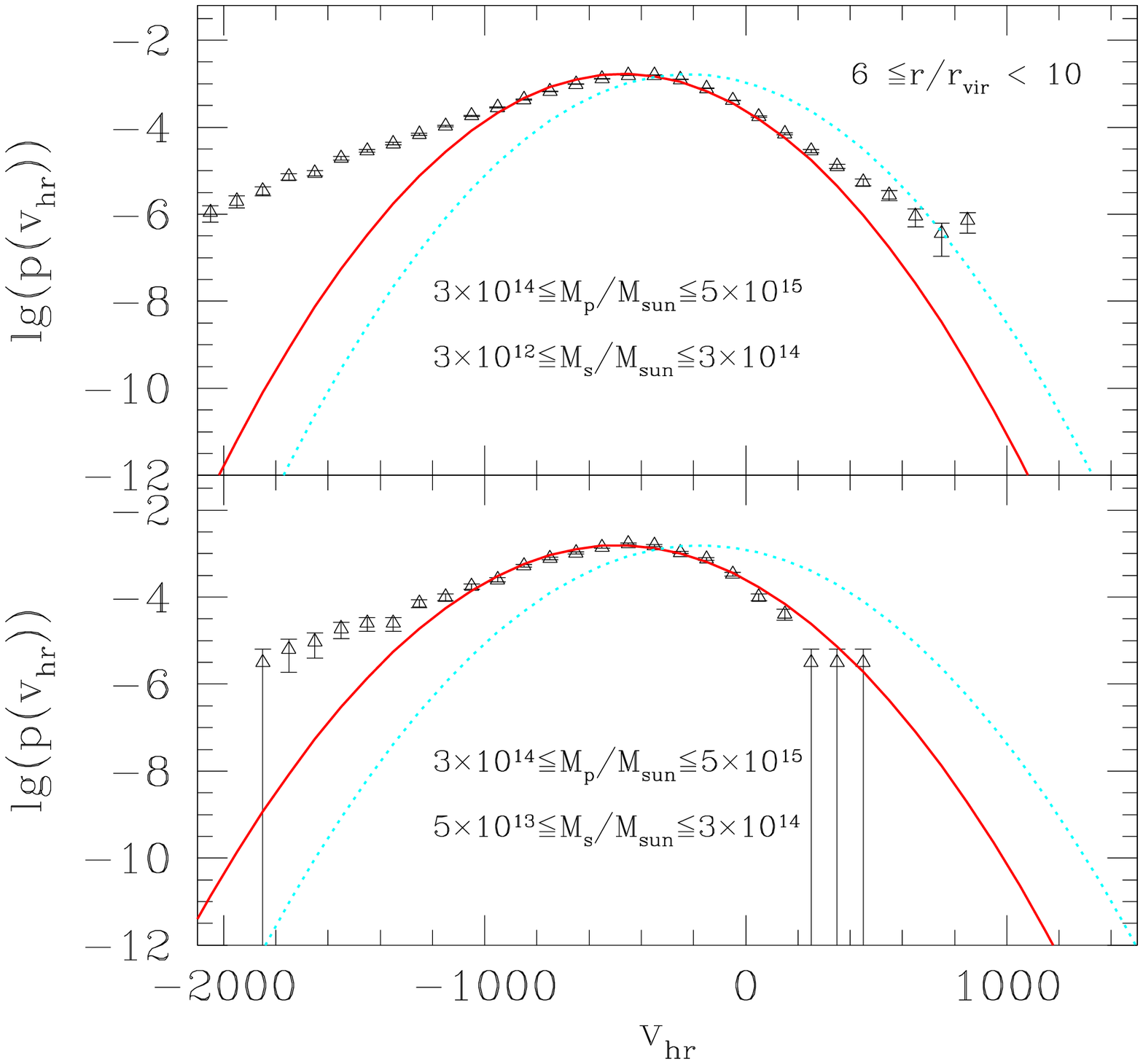}\\
\includegraphics[width=0.4\textwidth]{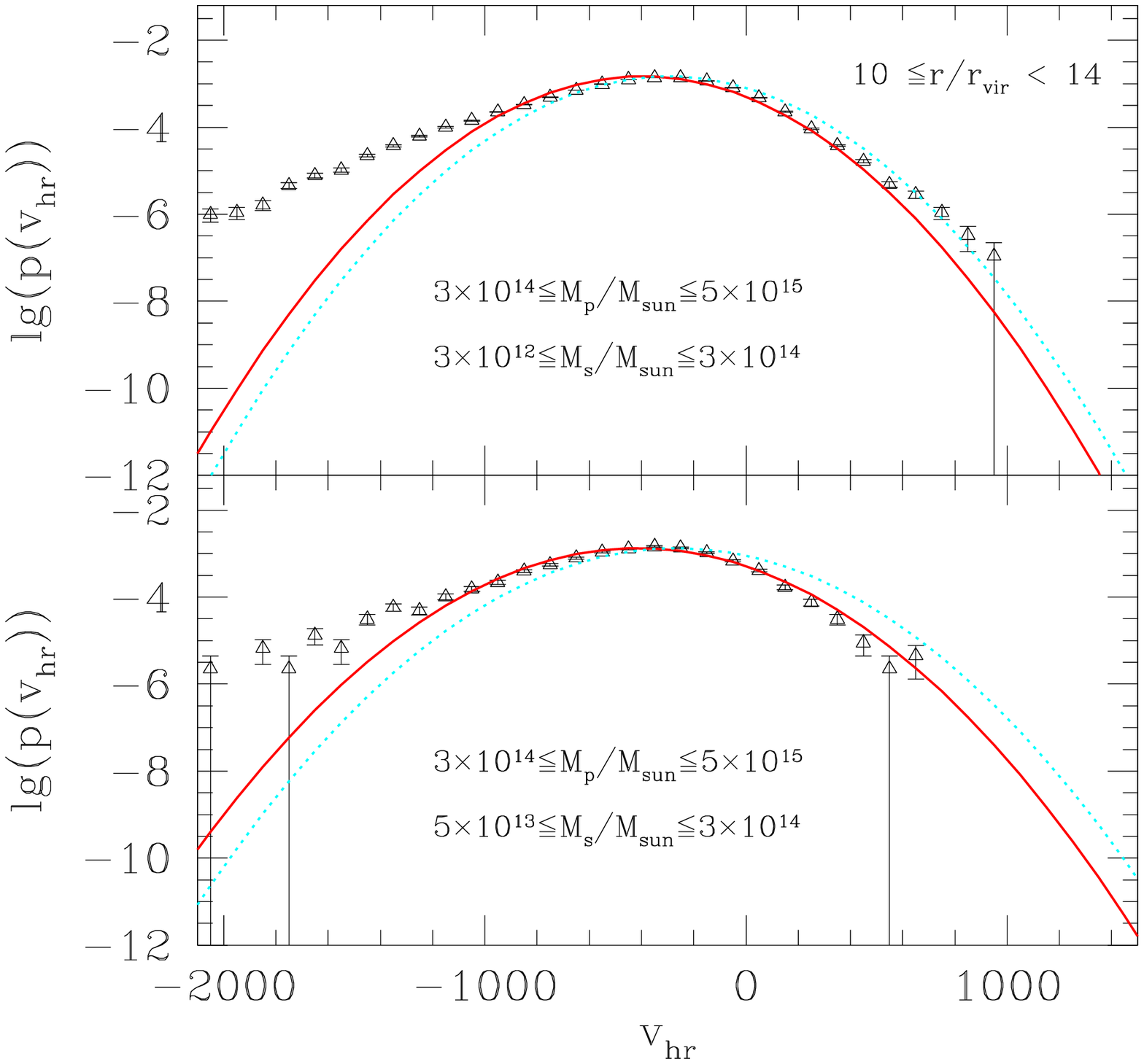}
\caption{The distribution of radial component of the halo-halo
  velocity. Pairs of halos whose separation are within $(6,10)$ and
  $(10,14)$ in the unit of $r/r_{\rm vir}$ are included in the
top and bottom panels respectively.}
\label{fig:pvll}
\end{figure}

While our model matches the lowest moments of the radial pairwise
velocity, we have not yet tested whether the assumption of a Gaussian
distribution is justified.  
\reffig{pvll} shows the radial velocity distribution of halos at two
separations ($r/r_{\rm vir}= 8$ on the top and $12$ on the
bottom respectively). 
The primary and secondary halo mass range matches one of the sets shown in 
\reffig{vhr}.
The cyan and red curves show the
corresponding linear theory and our empirical model predictions --
both of them are Gaussian distributed with the same variance but they
have different means. 
While the measured distributions are skewed towards
infalling radial velocity (negative velocity) and have exponential
wings, 
the empirical model is
able to match the peak regions of the distributions.  Thus, employing
an exponential rather than Gaussian model would improve the fit to
this distribution.  However, from our analysis of the contributions from the different components in 
the line-of-sight velocity dispersion described in the main text,
having an accurate description of the mean radial infall is the 
most important requirement to match the phase-space distribution,
and our empirical model is able to reach good agreement.

\begin{figure}
\centering
\includegraphics[width=0.5\textwidth]{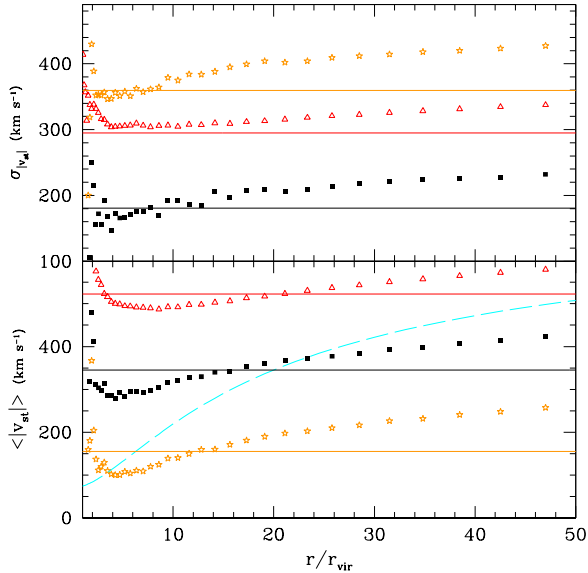}
\caption{The tangential component of the halo-halo pairwise velocity
         where $|v_{st}| = \sqrt{v_{ht,a}^2 + v_{ht,b}^2}$ while $v_{ht,a}$
         and $v_{ht,b}$ are  two tangential components.
         (see \reffig{vhr} for mass ranges used). 
        The mean and the dispersion are shown respectively in the 
        bottom and the top panels. Solid curves are the predictions
        of our empirical model, which assumes a constant given by the linear theory prediction at
        $r/r_{\rm vir}=20$. The red and orange sets are displaced by 
        $\pm 150 {\rm km/s}$. 
        The cyan dash curve in the 
        bottom panel shows the linear theory prediction for mass range 
        represented by the solid black squares.
        The red and the orange sets in the upper panel are 
        displaced by 100 km/s and 200 km/s respectively. 
}
\label{fig:vht}
\end{figure}

\reffig{vht} shows the mean and the dispersion of the 
tangential component of the halo pairwise velocity where we define
\begin{equation}
|v_{st}| = \sqrt{v_{ht,a}^2 + v_{ht,b}^2}.
\end{equation}
As described in the main text we will choose a projection of $|v_{st}|$ 
when evaluating the line-of-sight component. However in this section
we will perform the comparison using $|v_{st}|$ since
it is more straightforward to measure  from numerical 
simulations.
In contrast to the radial component, 
the mean and the dispersion of the tangential component do not 
have a strong scale dependence.
Notice that the spherical collapse model, by definition, does not 
predict any tangential component.
The linear theory prediction on the mean of the tangential component
is shown by the cyan dashed curve in the bottom panel (for the mass range
represented by the solid black squares). 
The linear theory does not match the measurements on all scales.  Hence, 
we make use of the apparent weak scale dependence 
of both the mean and the dispersion of the tangential component,
and take the linear theory prediction for the mean at 
$r = 20 r_{\rm vir}$ as our model.  Further, we assume $|v_{st}|$
is Rayleigh distributed with parameters
\begin{equation}
\langle |v_{st}| \rangle = \sqrt{\frac{\pi}{2}}\sigma_{ht} \quad {\rm and}
\quad \sigma_{|v_{st}|}^2 = \frac{4-\pi}{2}\sigma_{ht}^2,
\label{eqn:rayleigh}
\end{equation}
where $\sigma_{ht}$ is the dispersion which in our model is given
by \refeq{siguht} with $r=20 r_{\rm vir}$.  In other words,
we assume that each component $v_{ht,a}$ follows a Gaussian with
variance $\sigma_{ht}^2$.  
The solid curves in the upper panel of \reffig{vht} 
show the predictions of the empirical model using 
the relation in \refeq{rayleigh} -- at small to intermediate
separation (a few times to 20 times the virial radii) the 
predictions agree with the measurements at around 10\% level.

\begin{figure}
\centering
\includegraphics[width=0.4\textwidth]{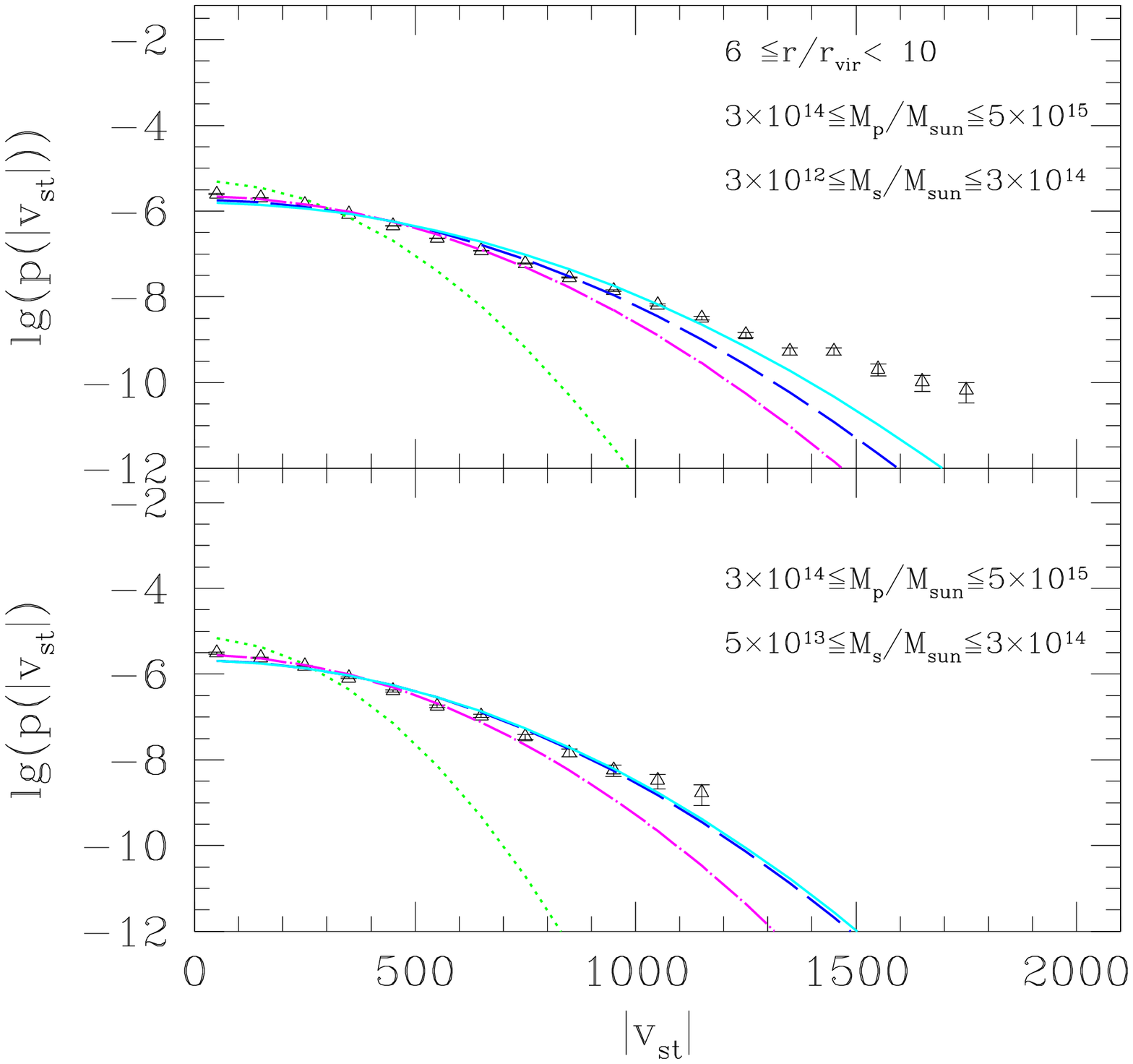}
\includegraphics[width=0.4\textwidth]{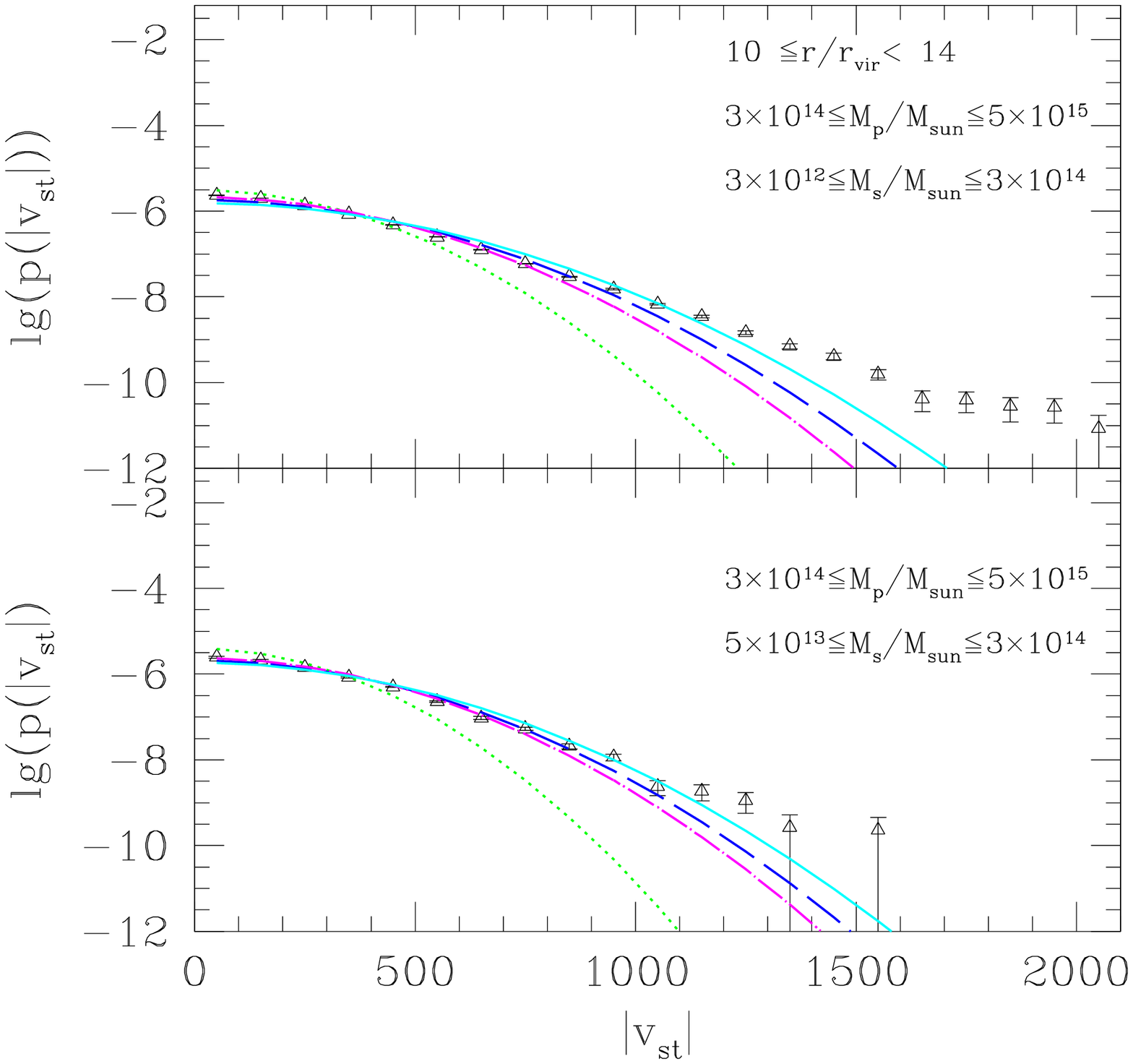}
\caption{The distribution of tangential component of the halo-halo
  velocity. The mass ranges of the primary and secondary halos 
   are the same as \reffig{pvll}. All curves are Rayleigh
   distribution (with proper normalization, see main text) with
   different parameter $\sigma$ (\refeq{rayleigh}): 
   cyan (solid) curves use the measured
   variance from simulation; green (dotted) curves use linear theory expected
   value at $r$; magenta (dot-dashed) use the measured mean from the
   numerical measurement; blue (dashed) use linear theory mean at a
   single separation $r/r_{\rm vir}= 20$.}
\label{fig:pvp}
\end{figure}

\reffig{pvp} shows the tangential velocity distribution at two
different separations. The separations and the halo mass ranges
are the same as in \reffig{pvll}.  Note that our
normalization requires that $2\pi\int {\rm d}|v_{st}| |v_{st}| p(|v_{st}|) = 1$.
By assuming the Rayleigh distribution we compare various 
ways to obtain the parameter $\sigma_{ht}$ at the corresponding separation:
cyan (solid) curves use the measured variance $\sigma^2_{|v_{st}|}$ 
from the simulation; 
magenta (dot-dashed) curves use the measured mean $\langle |v_{st}|\rangle$;
green (dotted) curves use the linear theory prediction.
The blue (dashed) curves are the empirical predictions in which we use
the linear theory predicted mean at one \emph{single} 
separation $r/r_{\rm vir}= 20$. 
Similar to the radial velocity, \reffig{pvp} shows that the distribution of the tangential velocity does not follow a
Rayleigh distribution, but is closer to an exponential.  
\citet{lnyfnlvel} describes an analytical model based on Zel'dovich
approximation that would mix initial radial component into the evolved
tangential component.
While the distribution profile from the measured variance (cyan solid curves)
match the measurements better at high tangential velocity,
we decide to adapt 
the empirical model using the linear theory prediction at
a \emph{single} separation $r/r_{\rm vir}=20$: this  
can be computed analytically from the linear power spectrum (hence can be 
extended to modified gravity models easily),  and it provides a reasonable
match to the measured distribution.

One important observation from \reffig{vhr} and \reffig{vht} is that
the mean and dispersion of the pairwise halo velocity depends only
weakly on the mass range of the secondary halos.  
The only exception is in the
tangential velocity distribution at small scales.  
The changes in the mean values of the velocity components are small
over the mass ranges considered: 
around 10\% for the radial component, and around 15-30\% for the
tangential component when the separation is small (a few times the
virial radius of the primary halo).
Similarly,  decreasing the lower bound for the primary halo mass does 
not change the halo pairwise velocity statistics significantly, 
where the most noticeable change is in the mean of the radial velocity 
at small separations. 

One aspect that is missing in the empirical model is the correlation
between the radial and the tangential velocity components. 
We have checked that while correlations between the two components 
do exist, they are weak (correlation coefficient $<0.3$). We also checked 
the conditional distribution of $v_{hr}$ and $|v_{st}|$ and 
found that only the dispersion of the radial component $\sigma_{hr}$ 
depends very weakly on $|v_{st}|$.  We expect that including these correlations
would not significantly impact our model predictions.


\section{Evaluation of the line-of-sight velocity dispersion}\label{sec:Qvir}
We are interested in evaluating the line-of-sight velocity dispersion 
of the phase-space distribution using the characteristic function. 
In this appendix we 
 illustrate this calculation using the 2-halo term of the  
phase-space distribution from halo-dark matter pairs as example, as 
the 1-halo term and the halo-halo pair cases are merely simpler versions of this
calculation. 
The characteristic function is the Fourier transform of the probability 
density function and provides an easy approach to compute 
all moments by taking derivatives with respect to the Fourier space
counterpart to $v_{\rm los}$, which we will denote as $t_{\rm los}$.

The corresponding characteristic function of $p(v_{\rm los}|r_p)$ is
\begin{align}
& \mathcal{M}_{{\rm proj},2h}(t_{\rm los}|r_p) = \int\! {\rm d} v_{\rm los} e^{i t_{\rm los} v_{\rm los}} p(v_{\rm los}| r_p)
\vs
& = \frac1{\N'(r_p)} \int {\rm d}M_p n(M_p) \!\! \int\! {\rm d}z \!\int\! {\rm d}M_s n(M_s) 
\vs
& \qquad\times
  \rho_{\rm DM}(r(z,r_p) | M_p) \mathcal{M}_{\rm 2h}(t_{\rm los}| M_p, M_s),
\label{eqn:mz}
\end{align}
where $\mathcal{M}_{\rm 2h}(t_{\rm los})$ is the normalized characteristic function 
of $\vec{v}_{\rm halo}$ and $\vec{v}_{\rm vir,s}$ projected along
the line-of-sight direction: 
\begin{align}
\mathcal{M}_{2h}(t_{\rm los}|& M_p, M_s) 
 = \int{\rm d}\vec{v}_{\rm halo} \int {\rm d}\vec{v}_{\rm vir,s} \vs
& \times
   e^{it_{\rm los}(\vec{v}_{\rm halo} + \vec{v}_{\rm vir})\cdot \hat{z}}
     p_{2h}(\vec{v}_{\rm halo},\vec{v}_{\rm vir,s}|r,M_p,M_s).
\nonumber
\end{align}
For the moment we write the full vector of the virial velocity 
$\vec{v}_{\rm vir}$.

The square of the line-of-sight velocity dispersion $\sigma_{v_{\rm los}}^2$ 
is given by $-\M''(t_{\rm los}=0)$. 
Since the only $t_{\rm los}$-dependent term in \refeq{mz} is 
$\mathcal{M}_{\rm 2h}$, 
the velocity dispersion is simply a weighted sum of 
the second order moments of the joint distribution 
$p_{2h}(\vec{v}_{\rm halo},v_{\rm vir,s})$.
In general there are  both auto as well as cross
correlation terms among different velocity components. 
As described in the main text, we make several simplifying 
approximations for computational convenience, which then lead to 
\begin{align}
& \mathcal{M}_{{\rm proj}, 2h}(t_{\rm los}|r_p) = \vs
&       \frac{\int{\rm d}M_p\ n(M_p)\int {\rm d}z\ 
              \bar{\rho}_m [1+\xi_{h\delta}(r|M_p)]
                \mathcal{M}_{\rm halo}(t_{\rm los}) 
           \mathcal{Q}_{\rm vir}(t_{\rm los})}{\int{\rm d}M_p\ n(M_p)\int {\rm d}z\ 
             \bar{\rho}_m  [1+\xi_{h\delta}(r|M_p)]},
\nonumber
\end{align}
where  $\mathcal{M}_{\rm halo}(t_{\rm los})$ denotes the characteristic function
for the halo pairwise velocity, $\mathcal{Q}_{\rm vir}$ is given by
\be
\mathcal{Q}_{\rm vir}(t_{\rm los}) = \bar{\rho}_m^{-1}\int{\rm d}M_s\ n(M_s)M_s
               \mathcal{M}_{\rm vir,s}(t_{\rm los}|M_s), 
\label{eqn:qqvir}
\ee
and  $\mathcal{M}_{\rm vir,s}(t_{\rm los}|M_s)$ denotes the virial 
motion within secondary halos of mass $M_s$ which is Gaussian distributed.
The evaluation of $\mathcal{Q}_{\rm vir}$ depends on the halo mass function 
as well as the linear power spectrum and the scaling of $\sigma_{\rm DM}(M)$ with $M$. 
We derive the expression  in the next subsection;
the required expression is given 
by \refeq{Qvariance}. Notice that in the following 
we replace $\sigma_*^2$, the virial velocity 
dispersion for $M_*$ halo, in 
\refeq{Qvariance} by $\sigma_{\rm eff}^2$ in 
\refeq{sigeff1}.

The quantity $\mathcal{M}_{\rm halo}(t_{\rm los})$ describes the 
halo pairwise velocity distribution, projected along 
the line-of-sight direction. We use the radial velocity $v_{hr}$ and the 
tangential velocity $v_{ht}$ to parametrize the expression:
\begin{widetext}
\begin{align}
\mathcal{M}_{\rm halo}(t_{\rm los}) &= 
           \int{\rm d}v_{hr} {\rm d}v_{ht}\ p_{hh}(v_{hr},v_{ht}|r,M_p) 
      \exp\left[it_{\rm los}(v_{hr}\cos\phi + v_{ht}\sin\phi)\right] \nonumber \\*
     & = \exp\left[it_{\rm los}\langle v_{hr}\rangle \cos\phi 
           - \frac{\sigma_{hr}^2}{2}t_{\rm los}^2\cos^2\phi 
           -\frac{\sigma_{ht}^2}{2}t_{\rm los}^2\sin^2\phi   
     - \langle v_{hr}v_{ht} \rangle t_{\rm los}^2 \cos\phi\sin\phi  + \ldots\right]\,,
\end{align} 
\end{widetext}
where we have used the fact that $\langle v_{ht} \rangle$ vanishes.  
Note that following assumption 2 in the main text (see section~\ref{sec:2-halo}), we neglect the dependence of $p_{hh}$ on $M_s$.
The contributions to $\sigma_{v_{\rm los}}^2$ are 
$(\sigma_{hr}^2 + \langle v_{hr} \rangle^2) \cos^2\phi$ and 
$\sigma_{ht}^2 \sin^2\phi$ while the cross term vanishes after integrating 
over the line-of-sight direction (recall that $\cos\phi \propto z$).  
These are given in \refeqs{vrvar}, \eqref{eqn:vtvar}, and \eqref{eqn:vrmean}.

By symmetry, the first moments of $v_{\rm los}$ of all involved distributions
vanish $\mathcal{M}'(t=0) = \mathcal{Q}'(t=0)=0$.  Thus,
\begin{align}
& \left[\mathcal{M}_{\rm halo}(t_{\rm los}) \mathcal{Q}_{\rm vir}(t_{\rm los})\right]''|_{t_{\rm los}=0} = \vs
&\quad \mathcal{M}_{\rm halo}''(t_{\rm los}=0) + \mathcal{Q}_{\rm vir}''(t_{\rm los}=0)\,.
\end{align}
This leads to the expression given in \refeq{DMdisp}.

\subsection{The characteristic function of the virial velocity in 2-halo regime}
In this subsection we compute the contribution from the virial motion 
within secondary halos to the dark matter phase-space distribution.  
Note that this contribution is absent for halo tracers.  
The equation for this contribution is 
\begin{widetext}
\begin{equation}
\mathcal{Q}_{\rm vir}(t) = \int^{M_{s,{\rm max}}}_{M_{s,{\rm min}}}{\rm d}M_s \ {\rm d}^3 \vec{y}\ 
          \frac{\rho_{\rm NFW}(y;M_s) n(M_s)}{\bar{\rho}_m^*(r,M_p)}[1+\xi_{hh}(\vec{r}+\vec{y},M_p,M_s)] \M_{\rm vir}(t;M_s),
\label{eqn:Mvirgen}
\end{equation}
\end{widetext}
where $\bar{\rho}_m^*(r,M_p)$ denotes the normalization in the above equation for 
$t=0$.
$\mathcal{Q}_{\rm vir}$ describes the mass weighted characteristic function of the virial motion 
due to secondary halos. 
Attentive readers may notice that \refeq{Mvirgen} is 
different from \refeq{qqvir} -- we will explain in the following 
how \refeq{qqvir} is an approximation of \refeq{Mvirgen}.
We would like to remind that the above equation assumes the first two approximations
described in the main text: the virial velocity within the secondary 
halo is independent of the peculiar motion of the hosting halo; and  the
halo-halo pairwise velocity distribution is weakly dependent on the
mass of the secondary halo.

The above integration can  be computed numerically, in particular 
we compute the line-of-sight velocity dispersion numerically in 
\refapp{effsigma}.  We now make further approximations as listed
in \refsec{model}.  Specifically, we assume 
$\vec{r} + \vec{y} \approx \vec{r}$.
The above equation is then simplified to
\begin{widetext}
\be
\mathcal{Q}_{\rm vir}(t;r,M_p) = \int^{M_{s,{\rm max}}}_{M_{s,{\rm min}}}{\rm d}M_s \ 
                           \frac{M_s n(M_s)}{\tilde{\rho}(r,M_p)}[1+\xi_{hh}(r,M_p,M_s)] \M_{\rm vir}(t;M_s) \label{eqn:funR}
\ee
\end{widetext}
where $\tilde{\rho}(r,M_p)$ denotes the normalization by
setting $t=0$. Notice that these normalizations 
represent the mean density -- 
$\tilde{\rho}$ and $\bar{\rho}_m^*$ are not equal to 
$\bar{\rho}_m$ since they takes into account 
the clustering and hence the dependence on $r$ and $M_p$.
\refeq{funR} can be evaluated analytically 
 when the following assumptions are made \citep[see, for example,][]{haloreview}
\begin{enumerate}
 \item  $\sigma_{\rm DM}(M)$ scales as $M^{1/3}$; 
 \item  The power spectrum of the matter density contrast is a power-law
        $\propto k^{-1}$. 
\end{enumerate}
The resulting analytical expression for the 
Sheth-Tormen (ST \cite{st02}) mass function is given by
\begin{widetext}
\begin{eqnarray}
 \mathcal{Q}_{\rm vir}^{\rm ST} & = & \frac{A(p)}{1+b(M_p)\xi_{\delta\delta}(r)} \left\{\frac{1+b(M_p)\xi_{\delta\delta}(r)(1-1/\delta_c)}{(1+\sigma_*^2t^2/q)^{1/2}}  + \frac{b(M_p)\xi_{\delta\delta}(r)/\delta_c}{(1+\sigma_*^2t^2/q)^{3/2}} +b(M_p)\frac{\xi_{\delta\delta}(r)}{\delta_c}\left(1+\frac{\sigma_*^2}{q}t^2\right)^{p-3/2}\frac{\Gamma(3/2-p)}{2^{p-1}\sqrt{\pi}}\right. \nonumber\\*
 &&\qquad + \left.\left[1+b(M_p)\xi_{\delta\delta}(r)\left(1-\frac{1}{\delta_c}\right) + 2p b(M_p)\frac{\xi_{\delta\delta}(r)}{\delta_c}\right] \left(1+\frac{\sigma_*^2}{q}t^2\right)^{p-1/2} \frac{\Gamma(1/2-p)}{2^p\sqrt{\pi}} \right\}, \label{eqn:Rst}
\end{eqnarray}
\end{widetext}
where $\sigma^2_{*} = \sigma^2_{\rm vir}(M_*)$ denotes the virial velocity 
variance for a $M_*$ halo and $A(p)=[1+\Gamma(1/2-p)/2^p\sqrt{\pi}]^{-1}$ is 
the mass function normalization.  In our numerical calculations, we assume
values for the Sheth-Tormen mass function parameters of $p=0.3$, and $q=0.75$, 
The configuration space counterpart of 
$\mathcal{Q}_{\rm vir}^{\rm ST}$
involves a combination of $K_0$, $K_1$, $K_p$ and $K_{1-p}$ (the modified 
Bessel functions of various orders).  
Since $K_n(x) \propto x^{-1/2} \exp(-x)$ for
large $x$, these yield exponential wings of the
virial velocity distribution which are induced by the mass weighting.   
Note that these exponential tails are consistent with numerical 
measurements found in previous studies.

The corresponding variance is then given by the second derivative of the
characteristic function,
\begin{widetext}
\begin{align}
-\mathcal{Q}_{\rm vir}^{\rm ST ''} (t=0)& =
\frac{1}{1+b(M_p)\xi_{\delta\delta}(r)} \frac{\sigma^2_*}{q}
   \left[1+b(M_p)\xi_{\delta\delta}(r)\left(1+\frac{2}{\delta_c}\right)\right]\left[1-2A(p)p\frac{\Gamma
    (1/2-p) }{2^p\sqrt{\pi}}\right]\,. \label{eqn:Rvariance}
\end{align}
\end{widetext}

The model described in the main text makes another approximation by
setting $b(M_s) = 1$ -- this brings \refeq{funR} consistent with \refeq{qqvir}.
The effect on this approximation can be realized by setting 
the term $b(M_p)\xi_{\delta\delta}$ to zero
or taking the limit of $\delta_c$ to infinity
in \refeqs{Rst} and \eqref{eqn:Rvariance}:
\begin{align}
\mathcal{Q}_{\rm vir}^{\rm ST} =  A(p)
     \Bigg[& \left(1 + \frac{\sigma^2_{*}}{q}t^2\right)^{-1/2} \vs
& + 
         \frac{\Gamma(1/2-p)}{2^{p}\sqrt{\pi}}
                \left(1 + \frac{\sigma^2_{*}}{q}t^2\right)^{p-1/2}\Bigg],
\label{eqn:qvir}
\end{align} 
and
\begin{align}
-\mathcal{Q}_{\rm vir}^{\rm ST ''}(t=0) & =  \frac{\sigma^2_*}{q} \left[1 - 2A(p)p\frac{\Gamma(1/2-p)}{2^p \sqrt{\pi}}\right]
\label{eqn:Qvariance}
\end{align}

Note that in the above results $\sigma^2_* = \sigma^2_{\rm vir}(M_*)$ denotes
the virial velocity dispersion of an $M_*$ halo.  Thus, when applying
all approximations listed in \refsec{model} the virial velocity 
dispersion becomes independent of the primary mass.  We have found that
our model makes significantly more accurate predictions when retaining
the functional form of $\mathcal{Q}_{\rm vir}^{\rm ST}(t)$, but replacing
$\sigma_*$ with a dispersion $\sigma_{\rm eff}$ calculated with less
drastic approximations as described in the following section.

\subsection{Effective virial velocity dispersion $\sigma_{\rm eff}$ in 2-halo regime}
\label{sec:effsigma}

In the framework of the halo model, the virial motion within secondary 
halos is responsible for the broader velocity distribution of
dark matter tracers as compared to halo tracers (see the middle and 
bottom panels in \reffig{nhalo_1} at large $r_p$). 

The derivation in the previous section made several assumptions 
listed in \refsec{model} so that the analytical calculation 
of this contribution is tractable.  In this section we provide a
more accurate prescription for the variance entering in \refeq{Qvariance}
by dropping some of the more drastic assumptions. 

First, note that the only $t$-dependent term in \refeq{Mvirgen} is
$\mathcal{M}_{\rm vir}(t; M_s)$, which is the
variance of the virial motion within the secondary halo.   
Since $\mathcal{M}_{\rm vir}(t;M_s)$ is the characteristic function of
a Gaussian, and $-\mathcal{M}_{\rm vir}''(t=0; M_s) = \sigma_{\rm DM}^2(M_s)$ following
\refeq{sigM}, this becomes
\begin{align}
\mathcal{Q}_{\rm vir}''(t=0)& = \frac{1}{\bar{w}}\int {\rm d} M_s \int_0^{r_{\rm vir,s}} {\rm
  d}^3 y\ n(M_s) \nonumber \\*
 & \quad \times [1 + \xi_{hh}(|\vec{r}+\vec{y}|,M_p,M_s)] \vs
& \quad \times \rho_{\rm NFW}(y) \sigma_{\rm DM}^2(M_s),
\label{eqn:sigeff} \\
& \equiv \sigma_{\rm eff}^2(r,M_p)\,,\nonumber
\end{align}
where 
\begin{align}
\bar{w} & = \int {\rm d} M_s \int_0^{r_{\rm vir,s}} {\rm
  d}^3 y\ n(M_s) \nonumber \\*
 & \quad \times [1 + \xi_{hh}(|\vec{r}+\vec{y}|,M_p,M_s)] \rho_{\rm NFW}(y)
\end{align}
is the normalization factor, and 
the factor in the square brackets takes into account halo clustering.  
We assume an NFW profile $\rho_{\rm NFW}(y)$ \citep{nfw} truncated at the 
virial radius.  As throughout, the halo correlation function is approximated by a 
linear bias model with halo exclusion,
\begin{equation*}
\xi_{hh}(r,M_p,M_s) = \left\{
\begin{array}{ll}
b(M_p)b(M_s)\xi_{\delta \delta}(r), & r \geq r_{\rm
  vir,p}+ r_{\rm vir,s}\\
-1, & \text{otherwise, } 
\end{array} \right.
\end{equation*}
where $\xi_{\delta \delta}(r)$ is the matter correlation function.  

\begin{figure}
     \centering \includegraphics[width=.4\textwidth]{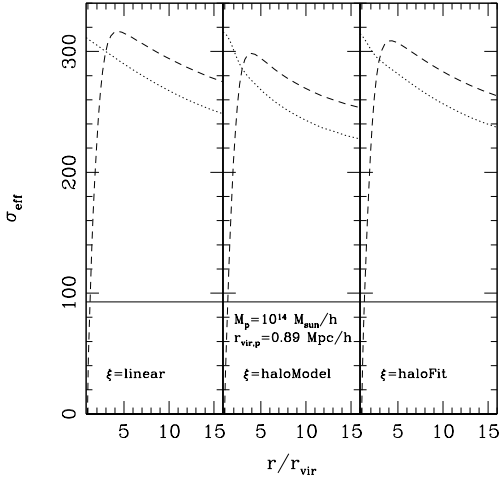}
  \caption{Direct numerical evaluation (dashed lines) of the line-of-sight projection 
           of the virial velocity dispersion within secondary halos
  [\refeq{sigeff}], for primary halos with $M_p = 10^{14} \Msunh$.  The panels show the result using different matter correlation functions 
  (linear correlation function on the left; halo model correlation
  function \citep{haloreview} in the middle; halofit correlation
  function \citep{smithetal} on the right).
  The dotted curves show the approximation in \refeq{sigeff1d} while the
  solid curves are the $M_*$ approximation $\sigma_{\rm eff} = \sigma_{\rm DM}(M_*)$.}
  \label{fig:sigeff}
\end{figure}

The model presented in the main text uses the full expression
\refeq{sigeff}.  However, since \refeq{sigeff} is a 3-dimensional integral (after making use of
the symmetry in the azimuthal angle), this calculation is
computationally demanding to perform for all possible pairs of $M_p$ and $r$.  We thus discuss possible approximations which speed up the 
calculation in the following.  
First, we neglect the halo exclusion effect, so that $\xi_{hh}(r)$ is simply a product of 
bias factors and the matter correlation function.  
Next, given that we are interested in scales $r_p$ much larger than
the virial radius of secondary halos which sets the typical value of 
$y$, we make the approximation that $|\vec{y}|=0$.
The above approximations simplify 3d integration into a 1d
integration over mass:
\begin{align}
& \int^{r_{\rm vir,s}}_0 {\rm d}^3y \
n(M_s)[1+\xi_{hh}(|\vec{r}+\vec{y}|,M_p,M_s)]  \rho_{\rm NFW}(y) \nonumber \\*
& \approx M_sn(M_s) [1 + b(M_p) b(M_s)
\xi_{\delta\delta}(r)].
\label{eqn:sigeff1d}
\end{align}
The normalization factor $\bar{w}$ is modified accordingly.

\begin{figure}[t]
\centering \includegraphics[width=0.48\textwidth]{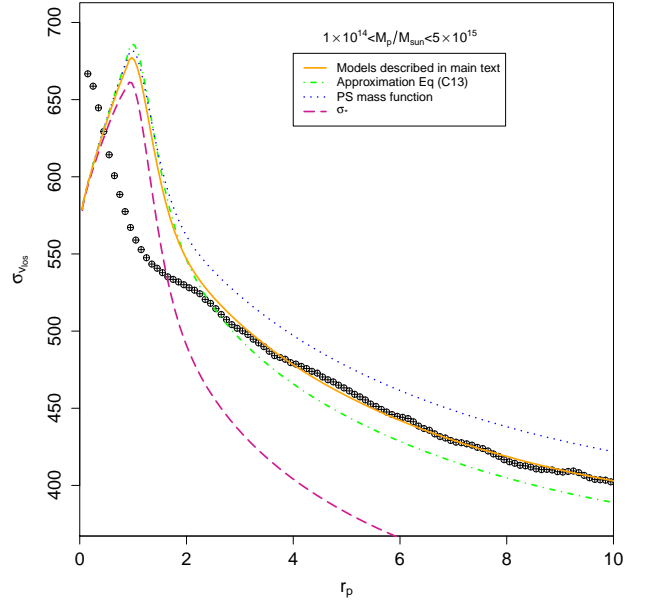}
\caption{Line-of-sight velocity dispersion as function of transverse separation, measured in simulations (black points) and predicted using different approximations for the secondary halo virial velocity treatment.
              Solid (orange) curves are 
              predictions from the model described in the main text 
              (same as \reffig{stats_1e14}).  Blue dotted curves use  
              the Press-Schechter mass function; green dot-dashed curves
               assumes $\vec{r}+\vec{y} \approx \vec{r}$ 
              (\refeq{sigeff1d}); 
                the magenta dashed curves
               use $\sigma_{\rm eff} = \sigma(M_*)$. 
   }
\label{fig:stats_1e14MANY}
\end{figure}

A further  approximation (``$M_*$ approximation'') can be made by assuming all the
secondary halos are $M_*$ halos where $M_*$ is the characteristic mass
scale such that $\sigma(M_*,z) = \delta_c$. 
It is equivalent to adding a delta function $\delta_{\rm D}(M_s -
M_*)$ to the above approximation and the resulting effective velocity
dispersion is
$\sigma^2_{\rm eff}(r,M_p) = \sigma_{\rm DM}^2(M_*)$ and it is
scale-independent.

\reffig{sigeff} (dashed lines) shows the result of the full expression \refeq{sigeff} as a
function of $r$, for primary halos
with $M_p = 10^{14} \Msunh$ and different matter correlation functions 
(see labels in the figure). 
The dotted and solid curves show the approximation in
\refeq{sigeff1d} and the $M_*$ approximation, respectively. 
The full calculation has a peak velocity dispersion at around a few
times the virial radius.  On smaller separation the halo exclusion
effect suppresses the velocity dispersion by excluding progressively
more massive halos.  
The approximation in \refeq{sigeff1d} traces the
scale-dependence of the full calculation well for $r \gtrsim 5 r_{\rm vir}$,
although there is a constant bias due to the change in the normalization
factor $\bar{w}$.  
The $M_*$ approximation gives a constant velocity dispersion which is
well below the full calculation.  

In order to assess the impact of these approximations on the
predicted phase-space distribution, we show the predictions for $\sigma_{v_{\rm los}}$
(for dark matter tracers) using different approximations in \reffig{stats_1e14MANY}.  The orange curve shows the default model discussed in the main text (using Sheth-Tormen mass function and \refeq{sigeff}).  The
blue (dotted) curves use the Press-Schechter mass function instead of 
Sheth-Tormen;
green (dot-dashed) curves assume $\vec{r}+\vec{y}\approx \vec{r}$ when 
evaluating the effective virial velocity dispersion 
(\refeq{sigeff1d});  
purple (long dashed) curves use $\sigma_*$ (``$M_*$ approximation''). 
Using the Press-Schechter mass function underestimates the 
line-of-sight velocity dispersion deep in the 2-halo regime.
Simplifying the calculation of the effective virial velocity by making
the assumption $\vec{r}+\vec{y}\approx \vec{r}$ reduces the 
computation time at the expense of 
underpredicting 
the measured 
velocity dispersion, consistent with \reffig{sigeff}.
Using $\sigma_*$ in the calculation yields a line-of-sight velocity
dispersion which is much too small, as expected from \reffig{sigeff}.

\end{document}